\documentclass[11pt, reqno]{article}
\usepackage{jheppub}
\usepackage{amssymb}
\usepackage{amsmath}
\usepackage{amsthm}
\usepackage[usenames,dvipsnames]{xcolor}
\usepackage{amsmath,amssymb,mathrsfs,amsthm,tikz,shuffle,paralist}
\usepackage{dsfont}
\usepackage{color}
\definecolor{darkred}{RGB}{173,34,48}
\usetikzlibrary{snakes}
\usetikzlibrary{calc}
\usetikzlibrary{decorations}
\usepackage[all]{xypic}
\usepackage{jheppub}
\usepackage{subfigure}
\usepackage{caption}
\usepackage{graphicx}            
\usepackage{extarrows}
\usepackage{amsmath}
\usepackage{mathtools}
\usepackage{epsfig}
\usepackage{tikz}
\usepackage{subfigure,tikz}
\usetikzlibrary{backgrounds,automata}
\usepackage{amssymb}
\usepackage{graphics}
\usepackage{subfigure}
\usepackage[active]{srcltx}
\usepackage{amsthm}
\usepackage[T1]{fontenc} 
\usepackage{graphicx}

\newcommand{\dd}{\text{d}}
\newcommand{\bea}{\begin{eqnarray}}
\newcommand{\eea}{\end{eqnarray}}
\newcommand{\bean}{\begin{eqnarray*}}
	\newcommand{\eean}{\end{eqnarray*}}
\newcommand{\nn}{\nonumber \\}

\def\co{\,,}
\def\ed{\,.}
\def\newline{{\hspace{15pt}}}
\def\no{\nonumber}

\def\abs#1{\left| #1\right|}

\def\det{\mathop{\rm det}}

\def\eref#1{(\ref{#1})}

\def\d{{\rm d}}

\def\a{{\alpha}}

\def\b{{\beta}}

\def\d{\partial}

\def\eps{\epsilon}

\def\what{\widehat}

\newcommand{\overtr}[1]{(\overline{#1})}
\newcommand{\adovertr}[1]{\dot{(\overline{#1})}}
\newcommand{\adtr}[1]{\dot{(\overline{#1})}}

\newcommand{\tensorc}[3]{C^{(#1)}_{{#2}\to {#2;\widehat{#3}}}}
\newcommand{\tensorselfc}[2]{C^{(#1)}_{{#2}\to {#2}}}

\newcommand{\measure}[2]{{\langle{#1}\mathrm{d}^{#2}{#1}\rangle}}
\def\Label#1{\label{#1}%
	\smash{\hbox to0pt{\raise1ex\hbox{\tiny[#1]}\hss}}}

\title{\boldmath Universal Treatment of Reduction for One-Loop Integrals in Projective Space}

\author[a,b,c,d]{Bo Feng,}
\author[a]{Jianyu Gong,}
\author[a]{Tingfei Li}
\affiliation[a]{Zhejiang Institute of Modern Physics, Zhejiang University, Hangzhou, 310027, P. R. China }
\affiliation[b]{Beijing Computational Science Research Center, Beijing 100084, China}
\affiliation[c]{Center of Mathematical Science, Zhejiang University, Hangzhou, 310027, P. R. China}
\affiliation[d]{Peng Huanwu Center for Fundamental Theory, Hefei, Anhui 230026, China}
\emailAdd{fengbo@zju.edu.cn}
\emailAdd{jianyu\_gong@zju.edu.cn}
\emailAdd{tfli@zju.edu.cn}

\abstract{Recently a nice work about the understanding of one-loop integrals has been done in \cite{Arkani-Hamed:2017ahv} using the tricks of the projective space language associated to their Feynman parametrization. We find this language is also very suitable to deal with the reduction problem of one-loop integrals with general tensor structures as well as propagators with arbitrary higher powers. In this paper, we show that how to combine  Feynman parametrization and embedding formalism to give a universal treatment of reductions for general one-loop integrals, even including the degenerated cases, such as the vanishing Gram determinant. Results from this method can be written in a compact and symmetric form.
	
}


\begin{document}
	\maketitle
	\flushbottom
\section{Introduction}
\label{sec:introduction}
    In recent years we have witnessed enormous progress in computing and understanding analytic structures of scattering amplitudes. At one-loop level, it is well-known that a general one-loop integrals in $D=4-2\eps$-dimension  can always be reduced to a linear combination of one-loop scalar integrals with no more than 5 external legs with   reduction coefficients as rational functions of the external variables~\cite{Brown:1952eu,Melrose:1965kb,Passarino:1978jh,tHooft:1978jhc,vanNeerven:1983vr, Stuart:1987tt,vanOldenborgh:1989wn,Bern:1992em,Bern:1993kr,Fleischer:1999hq,Binoth:1999sp,Denner:2002ii,Duplancic:2003tv,Denner:2005nn,Ellis:2007qk,Ossola_2007, Bern:1994cg}. These master integrals at one-loop level (i.e., tadpoles, bubbles, triangles, boxes, and pentagons) are well known. Therefore, the main problem of one-loop integrals is to calculate the reduction coefficients. There are a host of methods to deal with the reduction  at integrand level and integral level, such as Integration-By-Parts (IBP) \cite{Chetyrkin:1981qh,Tkachov:1981wb}, Passarino-Veltman reduction~\cite{Passarino:1978jh}, Ossola-Papadopoulos-Pittau reduction~\cite{Ossola:2006us,Ossola:2007bb,Ellis:2007br}, and unitarity cut ~\cite{Bern:1994zx,Bern:1994cg,Britto:2004nc,Britto:2005ha,Campbell:1996zw,Bern:1997sc,Denner:2005nn,Anastasiou:2006gt,Britto:2010um}. 
    
    Although in practice, we will not meet many situations where propagators have higher power, a complete reduction method should be able to deal with it. From this point of view, IBP method is a complete method, because it treats these complicated cases with the same framework as the one without higher poles.  Recently, combining unitarity cut and derivation over mass, the reduction coefficients for higher pole cases can be calculated~\cite{Feng:2021spv}, except for the tadpole coefficients. 
    
    Recently, we have proposed an improved PV-reduction method for one-loop integrals ~\cite{Feng:2021enk,Hu:2021nia}. The reduction coefficients can be expressed with the cofactors of the Gram matrices and have some symmetry. Thus it is useful to understand these symmetries appearing in our middle recursion relations and the final results. Notably, the analytical structure of one-loop integrals is studied by investigating Feynman parametrization in the projective space for its compactness and the close relation to geometry~\cite{Arkani-Hamed:2017ahv}. Inspired by the geometric angle, we find it could be convenient to do reduction for one-loop integrals in projective space. By our
    study in this paper,  one can see that the symmetry and simplicity of reduction coefficients are illustrated clearly with the denotations in~\cite{Arkani-Hamed:2017ahv}.
    
    Motivated by the work \cite{Arkani-Hamed:2017ahv}, we will develop an alternative method for determining the reduction coefficients of one-loop integrals in $D=4-2\eps$ dimension. The general tensor integrals with higher poles are related to integrals $E_{n,k}[T]$ in projective space
  \begin{align}
  E_{n,k}[T]\equiv \int_{\Delta}{\measure{X}{n-1}T[X^k]\over (XQX)^{n+k\over 2}}\co
  \end{align}
  where $\Delta$ is a simplex in $n$-dimensional space $X_I$, which is defined by $H_IX=X_I>0,\forall I=1,2,\ldots,n$. The $T$ is  a general tensor, which is contracted with $k$ $X$'s.
  
  The homogeneous coordinates $X_I$ are denoted by a square bracket $X=[x_1:x_2:\ldots:x_n]$, and two coordinates are  equivalent to each other up to a scaling, i.e., $[x_1:x_2:\ldots:x_n]\sim [kx_1:kx_2:\ldots:kx_n]$ for any $k\neq 0$.
  
  The measure in the projective space is given by the differential forms 
  
  \begin{align}
  \measure{X}{n-1}={\epsilon^{I_1,I_2,\ldots,I_n}\over (n-1)!}X_{I_1} \dd X_{I_2}\wedge \dd X_{I_3} \wedge \ldots \wedge \dd X_{I_n};~~
  XQX=Q^{IJ}X_IX_J\ed
  \end{align}
  
   As pointed out in \cite{Arkani-Hamed:2017ahv}, the integral $E_{n,k}$ 
   satisfies a nice recursion relation, which will be recalled in Appendix \ref{sec:proof of Enk}. It is exactly the property  we will use to do the reduction in this paper. 
   
   This paper is structured as follows. In section \ref{sec:integrals in projective}, we discuss how to write a general one-loop integral as the sum of integrals $E_{n,k}$ in projective space. In section \ref{sec:one-loop reduction}, we derive recursion relations for $E_{n,k}[V^i\otimes L^{k-i}]$ and dimension recursion relations for non-degenerate $Q$, and then apply them to the  reduction of one-loop integrals. In section \ref{sec:deQ-reduction}, we discuss the reduction framework for degenerate $Q$. In section \ref{sec:ngonself}, we show how to obtain the general expression of reduction coefficient from $n$-gon tensor integrals to $n$-gon scalar integrals while general expressions of reduction coefficients are given in Appendix \ref{apdix:generalcoeff}.  More reduction results are listed in Appendix \ref{sec:more results}.

\section{One-loop integrals in projective space}
\label{sec:integrals in projective}

In this paper, we will discuss the reduction of the most general one-loop integrals
\bea \int {d^D k\over i\pi^{D/2}}  {k^{\mu_1}k^{\mu_2}... k^{\mu_r}\over \prod_{j=1}^n D_j^{v_j}}=\int {d^D k\over i\pi^{D/2}} {k^{\mu_1}k^{\mu_2}... k^{\mu_r}\over \prod_{j=1}^n [(k-q_j)^2-m_j^2]^{v_j}}\co~~~~\label{TensorPole}
\eea
where $q_j=\sum_{i=1}^{j-1}p_i$.As pointed out in \cite{Feng:2021enk}, we can recover the tensor structure by multiplying each index with an auxiliary vector $R_{i,\mu_i}$. Furthermore, we can combine these $R_i$ to $R=\sum_{i=1}^r x_i R_i$ to simplify the expression \eref{TensorPole} in a Lorentzian invariant form to
\bea I^{(r)}_{\mathbf{v}_n;D}\equiv\int {d^D k\over i\pi^{D/2}}  {(2R\cdot k)^r\over \prod_{j=1}^n ((k-q_j)^2-m_j^2)^{v_j}}\ed~~~~\label{TensorPoleR}
\eea
To recover the result of \eref{TensorPole}, one can expand $R$ and extract the coefficient of $\prod_{i=1}^r x_i$ from the auxiliary formulas \eref{TensorPoleR}. Thus,   we will transform the general forms \eref{TensorPoleR} into projective space suggested in \cite{Arkani-Hamed:2017ahv}. First, to make our formulas elegant, we denote $y^{\mu}\equiv k^{\mu},y_i\equiv q_i$. Then \eref{TensorPoleR} becomes
\bea I^{(r)}_{\mathbf{v}_n;D}=\int {d^D y\over i\pi^{D/2}}  {(2R\cdot y)^r\over \prod_{j=1}^n ((y-y_i)^2-m_j^2)^{v_j}}\ed~~~~\label{TensorPoleREmbedding}
\eea
Then we put the whole formula into the embedding space with two higher dimensions with
\bea
	y^\mu\longmapsto Y^M=(Y^+,Y^-,Y^\mu)=(1,y^2,y^\mu)\co~~~~\label{set-2-1}
\eea
where we use the light-cone coordinates, i.e. the metric $\eta_{+-}=\eta_{-+}=-\frac{1}{2}$, $\eta_{\mu\nu}=diag(+,----)$  while all other entries vanish. For clarity, we will use the capital letters $M,N$ to denote the components of vectors in the embedding space, Greek letters $\mu,\nu$ to denote the components of Lorentzian vectors, and lower-case letters $i,j$ for the external legs. We will also use capital letters $Y,X$ simultaneously to denote vectors in the embedding space and projective space without ambiguity. Therefore, we can simplify the denominator of \eref{TensorPoleREmbedding} into the inner product of two vectors in the embedding space, and the quadratic expression has been somehow linearized, for example,
\bea 
	(y-y_i)^2=-2Y\cdot Y_i\ed~~~~\label{set-2-2}
\eea
After defining
\bea \mathcal{Y}_i^M=(1,y_i^2-m_i^2,y_i^\mu)\co~~~Y_\infty^M=(0,1,0,\ldots,0)\co~~~\mathcal{R}^M= (0,0,R^\mu)\co~~~~\label{set-2-3}\eea
it is easy to check that \eref{TensorPoleREmbedding} becomes
\bea I^{(r)}_{\mathbf{v}_n;D}=\int[\mathrm{d}^DY]\frac{(-2\,Y\cdot Y_\infty)^{v-D-r} (2Y\cdot \mathcal{R})^r}{\prod_{j=1}^n(-2\,Y\cdot \mathcal{Y}_j){}^{v_j}}\co~~~~\label{set-1-4}
\eea
where $v=\sum_{j=1}^n v_j$. The projective space invariant  measure is given by  $\int[\mathrm{d}^DY]\equiv\int\frac{\mathrm{d}^{D+2}Y\,\delta(Y^2)}{i\pi^{D/2}\text{vol.}\mathrm{GL}(1)}$ ($\mathrm{GL}(1)$ acts as an overall scaling of the $Y$ coordinates) and the factor $(Y\cdot Y_\infty)^{v-D-r}$ is necessary for the last expression to be genuinely an integral over the projective light cone. Using the most general Feynman parametrization
\bean {1\over A_1^{m_1} A_2^{m_2} ... A_n^{m_n}}={\Gamma(\sum_i m_i)\over \prod \Gamma(m_i)}\int_0^1 dx_1...dx_n \delta\bigg(\sum_i x_i-1\bigg) { \prod x_i^{m_i-1}\over (\sum_i x_i A_i)^{\sum_i m_i}}\co\eean
and putting the Feynman parameters into the projective space, \eref{set-1-4} becomes
\bea I^{(r)}_{\mathbf{v}_n;D}={\Gamma(v)\over \prod \Gamma(v_i)}\int_\Delta \measure{X}{n-1} X^{\mathbf{v}_n-1}\int[\mathrm{d}^DY]\frac{(-2\,Y\cdot Y_\infty)^{v-D-r} (2Y\cdot \mathcal{R})^r}{  (-2Y\cdot W)^v }\co~~~~\label{set-1-5}
\eea
where $X=[x_1:x_2:\ldots:x_n]$, which is a vector in a different projective (Feynman parametrization) space, $W=\sum_j x_j \mathcal{Y}_j$ and there is an additional factor $X^{\mathbf{v}_n-1}\equiv \prod_{i=1}^n x_i^{v_i-1}=\prod_{i=1}^n (H_iX)^{v_i-1}$.    Then the Feynman parametrization integral has been written into the compact form $\int_\Delta \measure{X}{n-1}$ (see \cite{Arkani-Hamed:2017ahv} for more details). We can further simplify \eref{set-1-5} using the common trick as
\bea I^{(r)}_{\mathbf{v}_n;D}&= & {\Gamma(v)\over \prod \Gamma(v_i)}\int_\Delta \measure{X}{n-1} X^{\mathbf{v}_n-1} {(-)^{v+D+r}\Gamma(D)\over \Gamma(v)}\left(\mathcal{R}^M\frac{\partial}{\partial W^M}\right)^{r}
\nn & &~~~~\times \left(Y_\infty^M\frac{\partial}{\partial W^M}\right)^{v-D-r}\int[\mathrm{d}^DY]\frac{1}{  (-2Y\cdot W)^D }\ed~~~~\label{set-1-6}
\eea
Up to now, the last integral in \eref{set-1-6} can be done easily. One way to solve it is to translate it back to the form \eref{TensorPoleREmbedding}, which is\footnote{We have used the fact $W_-=1$ by the Feynman parametrization. The integration result can be found in the formula (A.44) in the book
of Peskin and Schroeder \cite{Peskin:1995ev}.}
\bea \int {d^D y\over i\pi^{D/2}} {1\over (W_-(y-{w\over W_-})^2+W_+-{w^2\over W_-})^D}={ \Gamma({D/2})\over(-)^{D} \Gamma(D)}
(W\cdot W)^{-D/2}\co\eea
and we have 
\bea I^{(r)}_{\mathbf{v}_n;D}&= & {(-)^{v+r}\Gamma(D/2)\over \prod \Gamma(v_i)}\int_\Delta \measure{X}{n-1} X^{\mathbf{v}_n-1} \left(\mathcal{R}^M\frac{\partial}{\partial W^M}\right)^{r}
\left(Y_\infty^M\frac{\partial}{\partial W^M}\right)^{v-D-r}(W\cdot W)^{-D/2}\ed~~~~~~~~
\eea
The action of $\left(Y_\infty^M\frac{\partial}{\partial W^M}\right)$ can be done easily after using the fact $Y_\infty\cdot Y_\infty=0$
and we get
\bea I^{(r)}_{\mathbf{v}_n;D}&= & {(-)^{v+r}\Gamma(D/2)\over \prod \Gamma(v_i)}\int_\Delta \measure{X}{n-1} X^{\mathbf{v}_n-1} {\Gamma(v-{D\over 2}-r)\over \Gamma(D/2)}(LX)^{v-D-r} \nn & & ~~~~\left(\mathcal{R}^M\frac{\partial}{\partial W^M}\right)^{r}
(W\cdot W)^{-(v-D/2-r)}\co~~~~\label{set-1-7}
\eea
where we have written $(-2Y_\infty\cdot W)=\sum_{i} x_i\equiv L\cdot X$ with $L=[1:1:...:1]$\footnote{The reason that we can ignore the action of $\left(\mathcal{R}^M\frac{\partial}{\partial W^M}\right)$ on $(-2Y_\infty\cdot W)$ is because $Y_\infty\cdot \mathcal{R}=0$.}. The action of $\left(\mathcal{R}^M\frac{\partial}{\partial W^M}\right)$ is more complicated. By power counting, we have the general expansion 
\bea
(\mathcal{R}\cdot \partial_W)^r(W^2)^k
&=&\sum_{i=0}^{r}\mathscr{C}^{k}_{r,i}(W^2)^{k-{r+i\over 2}}(R^2)^{r-i\over 2}(\mathcal{R}\cdot W)^i\co
~~\label{tensor-exp}\eea
where $k$ can be an arbitrary number and $i$ has the same parity as $r$ due to the power of $R^2$ must be an integer. The expansion coefficients $\mathscr{C}^{k}_{r,i}$ are determined by initial conditions $\mathscr{C}^{k}_{0,i}=\delta_{i,0}, \mathscr{C}^{k}_{1,i}=2k \delta_{i,1}$ and the recursion relation
\bea
\mathscr{C}^{k}_{r+1,i}=(i+1)\mathscr{C}^k_{r,i+1}+(2k-r-i+1)\mathscr{C}^k_{r,i-1}\ed~~~\label{C-coeff}
\eea
From the recursion relation, one can solve $\mathscr{C}^k_{r,i}$ for general $r,i$ as
\bea
\mathscr{C}^k_{r,i}=\frac{2^r r! \Gamma\big(  {r-i+1\over 2}\big)}{\sqrt{\pi }
	i! (r-i)!}\prod_{j=1}^{{r+i\over 2}}(k+1-j)=\frac{2^r r!k! \Gamma\big(  {r-i+1\over 2}\big)}{\sqrt{\pi }
	i! (r-i)!(k-{r+i\over 2})!}\co~~~ {r-i\over 2}\in \mathbb{N}\ed
\eea
Plugging \eref{tensor-exp} into \eref{set-1-7} we get
\bea I^{(r)}_{\mathbf{v}_n;D}&= &{\Gamma(v-{D\over 2}-r)\over (-)^{v+r} \prod \Gamma(v_i)}\sum_{i=0}^{r}\mathscr{C}^{D/2+r-v}_{r,i} 
(R^2)^{r-i\over 2}\int_\Delta {\measure{X}{n-1} X^{\mathbf{v}_n-1}(\mathcal{R}\cdot W)^i(LX)^{v-D-r}\over (W\cdot W)^{v-{D+r-i\over 2}}}\ed ~~~~~~~~~\label{set-1-8}
\eea
Since the remaining integrals are over the projective space of Feynman parameters, we should rewrite the formula by
\bea W\cdot W &= & (\sum_{a=1}^n x_a \mathcal{Y}_a)\cdot (\sum_{b=1}^n x_b \mathcal{Y}_b)
=\sum_{a,b} x_a (\mathcal{Y}_a\cdot  \mathcal{Y}_b) x_b= XQX\co \nn
\mathcal{R}\cdot W &= & \sum_{b=1}^n x_b \mathcal{R}\cdot \mathcal{Y}_b=V\cdot X\co~~~V=[R\cdot q_1: R\cdot q_2:\ldots: R\cdot q_n]
\ed~\label{X-form-1}\eea
Then we get 
\bea I^{(r)}_{\mathbf{v}_n;D}&= &{\Gamma(v-{D\over 2}-r)\over (-)^{v+r}\prod \Gamma(v_i)}\sum_{i=0}^{r}\mathscr{C}^{D/2+r-v}_{r,i} 
(R^2)^{r-i\over 2}\int_\Delta {\measure{X}{n-1} X^{\mathbf{v}_n-1}(VX)^i(LX)^{v-D-r}\over (XQX)^{v-{D+r-i\over 2}}}\ed~ ~~~~~~~\label{set-1-9}
\eea
For general one loop integrals \eref{TensorPole}, one can calculate $Q$ as 
\bea Q_{ij}={1\over 2}\left(m_i^2+m_j^2-q_{ij}^2\right)\co~~~~q_{ij}=q_i-q_j.~~~
\label{Q-value}\eea
Now the expression is written as the integration over the $X$-projective space. Using the result of \cite{Arkani-Hamed:2017ahv}
\begin{align}
E_{n,k}[T]\equiv \int_{\Delta}{\measure{X}{n-1}T[X^k]\over (XQX)^{n+k\over 2}}\co
~~~~\label{Enk-def}
\end{align}
where $\Delta$ is a simplex in $n$-dimensional space defined by $H_iX=X_i>0,\forall i=1,2,\ldots,n$ and  $T$ is  a $k$-th tensor contracted with $k$ $X$'s, the general one loop integral in the projective space \eref{set-1-9} can be written as
\bea I^{(r)}_{\mathbf{v}_n;D}&= &{\Gamma(v-D/2-r)\over (-)^{v+r}\prod \Gamma(v_i)}\sum_{i=0}^{r}\mathscr{C}^{D/2+r-v}_{r,i} 
(R^2)^{r-i\over 2} E_{n,2v-n-D-r+i}[\otimes _jH_j^{v_j-1}\otimes V^i]\co~~ ~~~~~~~\label{set-1-10}
\eea
where for simplicity, we write  $E_{n,k}[V^a\otimes L^{k-a}]\equiv E_{n,k}[V^a]$ by neglecting the power of $L$.

For the later use, we need to do symmetrization for the tensor $\otimes _jH_j^{v_j-1}\otimes V^i$. To do so, we can use the same trick as used in \eref{TensorPole} and \eref{TensorPoleR}, i.e., $Z\equiv\sum_{i=1}^{v-n}z_iH_i$ and $S=tV+Z$. Thus \eref{set-1-10} can be obtained from
\bea {\Gamma(v-D/2-r)\over (-)^{v+r}\prod \Gamma(v_i)}\sum_{i=0}^{r}\mathscr{C}^{D/2+r-v}_{r,i} 
(R^2)^{r-i\over 2} E_{n,2v-n-D-r+i}[S^{v-n+i}]~~ ~~~~~~~\label{set-1-11}
\eea
after taking the coefficients of $t^iz^{\mathbf{v}_n-1}\equiv t^i\prod_{i=1}^n z_i^{v_i-1}$. Taking care of numerical factors, the final expression is 
\bea
I_{\mathbf{v}_n;D}^{(r)}&=&\sum_{i=0}^{r} {i!\Gamma(v-D/2-r)\over (-1)^{v+r}  (v-n+i)!}\mathscr{C}^{D/2+r-v}_{r,i}(R^2)^{r-i\over 2} E_{n,2v-n-D-r+i}[ S^{v-n+i}]\Big\vert_{t^iz^{\mathbf{v}_n-1}}\ed~~~~~~~\label{UniversalCoeff}
\eea
A special case of \eref{UniversalCoeff} is that for $\mathbf{v}_n=\mathbf{1}_n$ and $r=0$, we have
\bea
I_{n;D}&=& (-1)^n{\Gamma(n-D/2)}E_{n,n-D}[L^{n-D}]\co~~~\label{E-I}
\eea
where $I_{n;D}$ is the scalar integral of $n$ propagators in $D$ dimension.

Before ending this section, we want to point out in our discussion, for example in \eref{set-1-5}, that the power $v-D-r$ could be positive or negative with the arbitrary choice of $r$. Since we have kept the dimension $D$ arbitrary, we can take $D$ properly (even a negative number) to make $v-D-r$ a positive integer to make later discussion legitimate. At the end of reduction, we can analytically continue $D$ to the proper dimension. We have checked with several examples that such a continuation is allowed. 

In this paper, we mainly discuss Feynman integrals in $D=4-2\eps$-dimension space. At the one-loop level, the master integrals are related to $E_{n,n-D}[L^{n-D}],n=1,2,\ldots,5$. So the main task of one-loop integral reduction is to reduce general integral $E_{n,k}[S^a\otimes L^{k-a}]$ to the basis $E_{n\le 5,n-D}[L^{n-D}]$.

\section{Reduction for non-degenerate $Q$}
	\label{sec:one-loop reduction}
	
	Having transformed our problem \eref{TensorPole} to the form \eref{UniversalCoeff},
	in this section, we will show how to use the tricks of integrals in projective space (see \cite{Arkani-Hamed:2017ahv}) to generate  recursion relations of $E_{n,k}[V^a\otimes L^{k-a}]$ . By applying these recursion relations iteratively, one can reduce a general one-loop integral to the basis with coefficients written by elegant expressions. In other words, the PV-reduction can be done universally in the new projective space form. As we will point out, the reduction coefficients will have an interesting pattern other than the obvious  permutation symmetry\footnote{In the work~\cite{Feng:2021enk,Hu:2021nia}, one can see that the recursion relations, for example, the bubble tensors to bubble basis and the triangle tensor to triangle basis, are similar, except the boundary conditions. This similarity has given an explanation for the interesting pattern we observed in this paper.}. Moreover, the reduction process can be carried out in \textsf{Mathematica} automatically.
	
	\subsection{Recursion relation}
	In this subsection, we derive the recursion relations of $E_{n,k}[V^a\otimes L^{k-a}]$. We first consider the case $Q$ is non-degenerate. The key equation is following (see Eq.(4.2) in  \cite{Arkani-Hamed:2017ahv})
	\bea
	{\measure{X}{n-1}T[X^k]\over (XQX)^{n+k\over 2}}&=&{1\over n+k-2}\dd_X\left[{\langle (Q^{-1}T)[X^{k-1}]X\dd^{n-2}X\rangle \over (XQX)^{n+k-2\over 2} }\right]\nn
	&&+{k-1\over n+k-2}{\measure{X}{n-1}(tr_Q T)[X^{k-2}]\over (XQX)^{n+k-2\over 2}}\co~~~\label{Base-Identity}
	\eea
	where $\dd_X=\dd X^{I}{\d\over \d X^I}, tr_Q T=Q^{-1}_{I_1I_2}T^{I_1I_2\ldots I_k}$. The proof of the formula can be found in the Appendix \ref{sec:proof of Enk}.
	By integrating \eref{Base-Identity} we get
	\bea
	E_{n,k}[T]=\alpha_{n,k}E^{(b)}_{n-1,k-1}[(H_bQ^{-1}T)]+\beta_{n,k}E_{n,k-2}[tr_Q T]\co~~\label{Enk-SymTReduce}
	\eea
	where summing over $b$ is implicitly and to simplify our denotations, we have defined
	\bea
	\alpha_{n,k}&\equiv&{1\over n+k-2}\co~~~\beta_{n,k}\equiv{k-1\over n+k-2}\ed\eea
	Let us give a little explanation for the first term on the right-hand side of \eref{Enk-SymTReduce}. When
	integrating a total derivative term, we should choose a patch. For simplicity, we assume $X_i=1$. Then we get the contribution from the boundary 
	$X_b=0 $ and $X_b=+\infty$ for $b \neq i$. By the dimensional regularization, the term 
	with  $X_b=+\infty$ gives zero. For the term $X_b=0$, in $\langle (Q^{-1}T)[X^{k-1}]X\dd^{n-2}X\rangle$, only when the first index of $Q^{-1}$ takes
	the value $b$, the contribution is nonzero, which is equivalent to write as
	$(H_bQ^{-1}T)$.
	
	When we repeatedly do the recursion relation, there will be a set of $X_b$ setting to zero. Writing the index set as
	$\mathbf{b}_j\equiv \{b_1,b_2,\ldots,b_j\}$ with $1\leq b_{i}<b_{i+1}\leq n$,
	we define $X_{(\mathbf{b}_j)}$ to be the new vector in lower-dimensional projective space obtained by removing 
	these components belonging to the set 
	$\mathbf{b}_j$ from the original vector $X$. With this understanding, the meaning of 
	\bea
	E^{(\mathbf{b}_j)}_{n-j,k}[T]&\equiv& 
	\int_{\Delta_{\mathbf{b}_j}}{T[X^k_{(\mathbf{b}_j)}]\measure{X_{(\mathbf{b}_j)}}{n-1-j}\over (X_{(\mathbf{b}_j)}Q_{(\mathbf{b}_j)}X_{(\mathbf{b}_j)})^{{n-j+k\over 2}}} \label{sub-Enk}
	\eea
	is clear. The equation \eref{sub-Enk} represents the integral got by removing propagators belonging to $\mathbf{b}_j$.

	Now we consider the reduction of $E_{n,k}[T]$ with $T=V^i\otimes L^{k-i}$ as given in \eref{UniversalCoeff}. Since tensor $T$ is contracted with $k$ X's, we can symmetrize its last $k-1$ indices as below
	\bea
	V^i\otimes L^{k-i}\to V\otimes {1\over (k-1)!}\sum_{\sigma \in S_{k-1}}\sigma \left[V^{i-1}\otimes L^{k-i}\right]\co
	\eea
	where the $\sigma$ is the permutation acting on the tensor $V^{i-1}\otimes L^{k-i}$. By applying \eref{Base-Identity}, one has
	\bea
	E_{n,k}[V^{i+1}]
	=\alpha_{n,k}\left [\overtr{H_bV}E^{(b)}_{n-1,k-1}[V^{i}]+{i }\overtr{VV}E_{n,k-2}[V^{i-1}]+(k-i-1)\overtr{VL}E_{n,k-2}[V^i]\right]\co\nn ~~~\label{Enk-ViRecursion}
	\eea
	where for simplicity the $L$ tensor part has been omitted. To make our formula more compact, here we have defined $\overtr{AB}\equiv
	A Q^{-1} B$. Here we want to remark a subtle point. In \eref{Enk-ViRecursion},
	the $Q$ is $n\times n$  matrix as defined in $E_{n,k}$ and appears in front of $E^{(b)}_{n-1,k-1}$ . When we try to iteratively use 
	\eref{Enk-ViRecursion} for $E^{(b)}_{n-1,k-1}$, that $Q$ will become the 
	$(n-1)\times (n-1)$  matrix $Q_{(b)}$. Thus, for later convenience, we define
	$\overtr{AB}_{(\mathbf{a}_j)}=A_{(\mathbf{a}_j)}(Q_{(\mathbf{a}_j)})^{-1}B_{(\mathbf{a}_j)}$ where matrix $(Q_{(\mathbf{a}_j)})^{-1}$ is the  
	inverse of the matrix obtained by removing the rows and columns of the index set
	$ \mathbf{a}_j$ from the original matrix $Q$.

\begin{figure}
	\begin{center}
		\begin{tikzpicture}[scale=1.3]
		\newcommand{\rectangle}[4]{\draw[fill,black] (#1-0.5*#3,#2-0.5*#4)--(#1+0.5*#3,#2-0.5*#4)--(#1+0.5*#3,#2+0.5*#4)--(#1-0.5*#3,#2+0.5*#4)--cycle;}
		\def\size{0.08}
		\def\boxsize{0.07}
		\def\offset{0.08}
		\newcommand{\Di}[2]{
		\ifnum #2>1
		\draw[->,line width=2,cyan](#1+\offset,#2-\offset)[bend left]to(#1+\offset,#2-2+\offset);
			\fi
			\ifnum #2>0
			\ifnum #1>1
		\draw[->,line width=2,red](#1-\offset,#2-\offset)to(#1-1+\offset,#2-1+\offset);
		\fi
		\draw[->,line width=2,orange](#1,#2-\offset)to(#1,#2-1+\offset);
		\fi}
		\draw[line width=1.2,->](-1,0)--(6,0);
		\draw[line width=1.2,->](0,0)--(0,6);
		\draw[fill,green] (0,0) circle [radius=0.08];
		\foreach \x in {0,1,2,...,5}{
			\node[below] at (\x,-0.1) {\x};
			\ifnum \x>0
			\node[left] at (-0.1,\x) {\x};
			\fi
			\node[below] at (\x,-0.1) {\x};
			\foreach \y in {0,1,2,...,5}{
				\ifnum \x>0
			    \draw[fill,blue] (\x,\y) circle [radius=\size];
				\fi
				\ifnum \x=0
				\draw[fill,black] (\x,\y) circle [radius=\size];
				\fi
				\ifnum \x>\y
					\Di{\x}{\y};
			    \fi
			}
		}
		\Di{1}{1};
		\Di{2}{2};
		\Di{3}{1};
		\Di{3}{3};
		\Di{4}{2};
		\Di{4}{4};
		\Di{5}{1};
		\Di{5}{3};
		\Di{5}{5};
		\node[below] at (5.8,0) {$n$};
		\node[left] at (0,5.8) {$i$};
		\end{tikzpicture}
		
		\caption{The reduction process of $E_{n=5,k}[V^{i=5}]$, where the black points represent zero terms. The red, orange and cyan arrows represent the first, second and third terms respectively in \eref{Enk-ViRecursion}.}
		\label{fig:EnkVi-REduce}
	\end{center}
\end{figure}
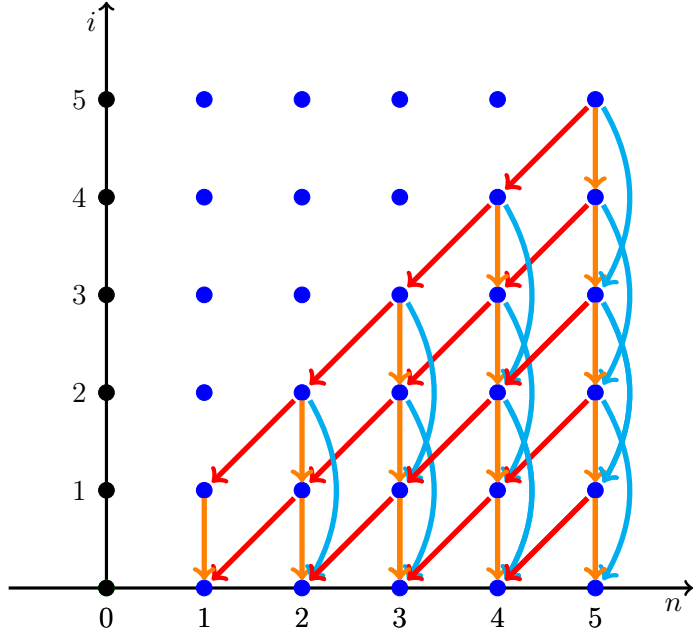

	As the main result of the whole paper, recursion relation \eref{Enk-ViRecursion} plays a crucial role in the reduction of one-loop integrals.  By comparing the power of $V$, one sees that it has been reduced from the LHS to RHS. Furthermore, from \eref{UniversalCoeff} one sees that the power is given by $v+i-n$, where $v$ contains the contribution of higher power of propagators and $i$ contains the contribution of the tensor numerator,  thus \eref{Enk-ViRecursion} provides the universal reduction of both cases.  As shown in the Fig\eref{fig:EnkVi-REduce}, after iteratively using
	\eref{Enk-ViRecursion}, we get a linear combination of $E^{(\mathbf{a}_{n-n'})}_{n'\leq n,k'<k}$, i.e., the scalar integral $I_{n';D'}$ in dimension $D'=n'-k'$. So starting with a general one-loop integral, one can always reduce it to the scalar integrals in different dimensions with coefficients being rational functions of external momenta. If we prefer the scalar basis in a given $D$-dimensional space, we need to find the formula to shift the dimension of the scalar basis to a fixed $D$.

	\subsection{Dimension recursion}
	As we have seen in \eref{E-I}, the integral $E_{n,k}$ corresponds to the scalar $n$-gon diagram in $(n-k)$-dimensional space. To find dimension recursion relations, we set $V=L$ in \eref{Enk-ViRecursion} and get
	\bea
	E_{n,k}=\alpha_{n,k}\overtr{H_bL}E^{(b)}_{n-1,k-1}+\beta_{n,k}\overtr{LL}E_{n,k-2}\ed~~~\label{Enk-V0Reduce}
	\eea
	To reduce $E_{n,k}$, where $n-k=D+2s,s\in \mathbb{Z},s\not=0$, we can iteratively use   \eref{Enk-V0Reduce}, which is established for the scalar integrals already. 
	Noticing that in the RHS of \eref{Enk-V0Reduce}, the first term has the same
	dimension as the LHS with one propagator removed, while the second term has two higher 
	dimensions with the same number of propagators. Depending on the sign of $s$, we can take different manipulations.  
	
	\begin{itemize} 
		\item $s>0$: For this case, we need to reduce an integral in a higher dimension to $D$-dimension,  so we solve the second term in the RHS of \eref{Enk-V0Reduce} and get
		\bea
		E_{n,k}={E_{n,k+2}-\alpha_{n,k+2}\overtr{H_jL}E^{(j)}_{n-1,k+1}\over \b_{n,k+2}\overtr{LL}}\ed\label{Dimension-Lower}
		\eea
		It is obviously that such a rewriting \eref{Dimension-Lower} is legitimate when and only when 
		$\overtr{LL}\not =0$. 
		For $\overtr{LL}=0$, we have
		\bea
		E_{n,k}=\alpha_{n,k}\overtr{H_jL}E_{n-1,k-1}^{(j)}\ed\label{Propagator-Cancel}
		\eea
		Both sides have the same dimension, but the  RHS of \eref{Propagator-Cancel} has one less propagator. One well-known example of $\overtr{LL}=0$ is that the bubble with null external momentum is not a basis anymore, and it is reduced to two tadpoles. 
		Having established \eref{Dimension-Lower} and \eref{Propagator-Cancel}, we can reduce $(D+2s)$-dimensional integrals to $D$-dimensional iteratively using either \eref{Dimension-Lower} or \eref{Propagator-Cancel} depending on  if  $\overtr{LL}$ is zero or not at that step\footnote{Please remember that as emphasized under \eref{Enk-ViRecursion}, at each step $Q$ is different.  }.

	    \item $s<0$: For this case, we  can use \eref{Enk-V0Reduce} directly.
	    \bea
	    E_{n,k}=\alpha_{n,k}\overtr{H_jL}E_{n-1,k-1}^{(j)}+\beta_{n,k}\overtr{LL}E_{n,k-2}\ed\label{Dimension-Higher}
	    \eea
	    As pointed out already, the first term on the RHS corresponds to the scalar integrals in the same dimension but with the $j$-th propagator removed  and the second term corresponds to scalar integrals in $D'=n-k+2$ dimension. For the boundary situation, i.e., $n=1$, the first term vanishes and the second term gives a higher dimensional scalar
	    basis. Repeating it, we can  reduce $E_{n,k}$ to scalar integrals in $D$-dimensional space.
	\end{itemize}

%
%
    \subsection{Examples}
    \label{sec:gQexamples}
     To illustrate our method and avoid complicated computation in general cases, we first consider the reduction of tensor bubbles
     \begin{align}
     I_2^{(r)}=\int {d^D\ell\over i\pi^{D/2}}{(2R\cdot \ell)^r\over D_1^{v_1}D_2^{v_2}}\co
     \end{align}
     where
     \begin{align}
     D_1=(\ell-q_1)^2-m_1^2\co~~~~ D_2=(\ell-q_2)^2-m_2^2\ed
     \end{align}
     We set $q_1=0 $ for simplicity, then
     \begin{align}
     L=\left[
     \begin{array}{c}
     1 \\
     1 \\
     \end{array}
     \right],
     H_1=\left[
     \begin{array}{c}
     1 \\
     0 \\
     \end{array}
     \right],H_2=\left[
     \begin{array}{c}
     0 \\
     1 \\
     \end{array}
     \right],Q=\left[
     \begin{array}{cc}
     m_1^2 & \frac{1}{2} \left(m_1^2+m_2^2-q_2^2\right) \\
     \frac{1}{2} \left(m_1^2+m_2^2-q_2^2\right) & m_2^2 \\
     \end{array}
     \right]\ed\label{3-13}
     \end{align}
     Here we give some results for different choices of the rank and the powers to illustrate our idea.
     \begin{itemize}
     \item \textbf{Tensor bubble with primary propagator} 
     	
     We first consider reducing a rank-1 bubble $I_2^{(1)}\equiv I_{\{1,1\}}^{(1)}$. Since there are no higher poles, we just choose $S=V,r=1,v=n=2$ in \eref{UniversalCoeff} and we have
     \begin{align}
     I_2^{(1)}=-{\Gamma(1-D/2)}\mathscr{C}^{D/2-1}_{1,1} E_{2,2-D}[V]=-{\Gamma(1-D/2)}(D-2) E_{2,2-D}[V]\ed\label{Bubbler1Exp1}
     \end{align}
     First, we use \eref{Enk-ViRecursion} to reduce $E_{2,2-D}[V]$ and get
     \bea
     E_{2,2-D}[V]=\a_{2,2-D}\overtr{H_iV}E^{(i)}_{1,1-D}+\b_{2,2-D}\overtr{VL}E_{2,-D}\co\label{Bubblev1Reduce}
     \eea
     where the first term $E^{(i)}_{1,1-D}$  corresponds to $D$-dimensional scalar tadpoles generated by removing the $i$-th propagator of bubble $I_2$ while the second term $E_{2,-D}$ corresponds to a $(D+2)$-dimensional bubble. We need to lower the dimension of the second term further. Here we assume $\overtr{LL}\not=0$, by using \eref{Dimension-Lower}, we get
     \begin{align}
     E_{2,-D}={E_{2,2-D}-\a_{2,2-D}\overtr{H_iL}E^{(i)}_{1,1-D}\over \b_{2,2-D}\overtr{LL}}\co\label{Bubbler1DimensionRecur}
     \end{align}
     where the two terms in the numerator correspond to the $D$-dimensional bubble and two tadpoles. Plugging \eref{Bubbler1DimensionRecur} and \eref{Bubblev1Reduce} to \eref{Bubbler1Exp1} and recognizing them as master integral according to \eref{E-I}  we have
     \begin{align}
     I_2^{(1)}=\sum_i-\frac{\overtr{L L} \overtr{H_i V}-\overtr{H_i L} \overtr{V L}}{\overtr{L L}}I_{2;\what{i}}
     + \frac{2 \overtr{V L}}{\overtr{L L}}I_2\ed
     \end{align}
     So the reduction coefficients are 
     \bea
     C^{(1)}_{2\to 2;\what{1}}&=&-\frac{\overtr{L L} \overtr{H_1 V}-\overtr{H_1 L} \overtr{V L}}{\overtr{L L}}=\frac{R\cdot q_2}{q_2^2}\co\nn
     C^{(1)}_{2\to 2;\what{2}}&=&-\frac{\overtr{L L} \overtr{H_2 V}-\overtr{H_2 L} \overtr{V L}}{\overtr{L L}}=-\frac{R\cdot q_2}{q_2^2}\co\nn
     C^{(1)}_{2\to 2}
     &=&\frac{2 \overtr{V L}}{\overtr{L L}}=\frac{\left(m_1^2-m_2^2+q_2^2\right) R\cdot q_2}{q_2^2}\ed\label{3-18}
     \eea
     \item \textbf{Tensor bubble with massless legs}
     
     One can notice there is a pole of $q_2^2$ in the reduction coefficients of $I_2^{(1)}$, which comes from $\overtr{LL}$ 
     \begin{align}
     \overtr{LL}=-\frac{4 q_2^2}{-2 \left(m_1^2+m_2^2\right) q_2^2+\left(m_1^2-m_2^2\right){}^2+q_2^4}\ed \label{3-19}
     \end{align}
     For $q_2^2=0$, we have $\overtr{LL}=0$, so we have
     \begin{align}
     E_{2,2-D}[V]=\a_{2,2-D}\overtr{H_bV}E^{(b)}_{1,1-D}+\b_{2,2-D}\overtr{VL}E_{2,-D}\ed
     \end{align}
     Here we need to reduce $E_{2,-D}$
     \begin{align}
     E_{2,-D}=E^{(b)}_{1,-D-1}={\a_{2,-D}\overtr{H_bL}E^{(b)}_{1,1-D}\over \b_{1,1-D}\overtr{LL}_{(b)}}\co\label{3-21}
     \end{align}
     where we have used \eref{Propagator-Cancel} and
     \begin{align}
      E^{(b)}_{1,-D-1}={E^{(b)}_{1,1-D}\over \b_{1,1-D}\overtr{LL}_{(b)}}\ed
     \end{align}
     Using \eref{E-I}, we finally get
     \begin{align}
     I_2^{(1);q_2^2=0}=\left[{1-D\over D}{\overtr{VL}\overtr{H_bL}\over \overtr{LL}_{(b)}}-\overtr{H_bV}\right]I_{2;\what{b}}\ed
     \end{align}
     Explicitly, we have
     \begin{align}
      I_2^{(1);q_2^2=0}=-\frac{2 \left(D m_1^2-(D-2) m_2^2\right) R\cdot q_2}{D \left(m_1^2-m_2^2\right)^2}I_{2;\what{1}}+\frac{4 m_1^2 R\cdot q_2}{D \left(m_1^2-m_2^2\right)^2}I_{2;\what{2}}\ed
     \end{align}
     \item \textbf{Scalar bubble with higher poles} 

     Then we consider reducing scalar bubbles $I_{\{v_1,v_2\}}$ with higher poles $v=v_1+v_2=3$. Due to there is no tensor structure in the numerator, we just set $S=Z=z_1H_1+z_2H_2, r=0$ in \eref{UniversalCoeff}
     \bea
     I_{\mathbf{v}_2;v=3}&=& {-\Gamma(3-D/2) } E_{2,4-D}[Z]\Big\vert_{z^{\mathbf{v}_2-1}}\co\label{3-25}
     \eea
     where $\mathbf{v}_2=\{1,2\},\{2,1\}$.  First, we use \eref{Enk-ViRecursion} to reduce $E_{2,4-D}[Z]$ and get
     \bea
     E_{2,4-D}[Z]=\a_{2,4-D}\overtr{H_iZ}E^{(i)}_{1,3-D}+\b_{2,4-D}\overtr{ZL}E_{2,2-D}\co\label{3-26}
     \eea
     where the first term $E^{(i)}_{1,3-D}$  corresponds to $(D-2)$-dimensional scalar tadpoles generated by removing the $i$-th propagator of bubble $I_2$ while the second term $E_{2,2-D}$ corresponds to wanted $D$-dimensional bubble. We need to lift the dimension of the first term further. By using \eref{Dimension-Higher}, we get
     \begin{align}
     E^{(i)}_{1,3-D}=\b_{1,3-D}\overtr{LL}E^{(i)}_{1,1-D}\co\label{3-27}
     \end{align}
     where we have used $E^{(ij)}_{0,2-D}=0$, and the RHS corresponds to $D$-dimensional scalar tadpoles. Plugging \eref{3-27} and \eref{3-26} to \eref{3-25} and recognizing them as master integrals according to \eref{E-I}  we have
     \begin{align}
     I_{\mathbf{v}_2;v=3}=-\frac{1}{4} (D-2) \overtr{L L}_{(i)} \overtr{H_i Z}\Big\vert_{z^{\mathbf{v}_2-1}}I_{2;\what{i}}+\frac{1}{2} (D-3) \overtr{Z L}\Big\vert_{z^{\mathbf{v}_2-1}}I_2\ed
     \end{align}
     There are two configurations  
     \begin{align}
     I_{\{2,1\}}&=-\frac{1}{4} (D-2) \overtr{L L}_{(i)} \overtr{H_i Z}\Big\vert_{z_1}I_{2;\what{i}}+\frac{1}{2} (D-3) \overtr{Z L}\Big\vert_{z_1}I_2\nn
     &=-\frac{1}{4} (D-2) \overtr{L L}_{(i)} \overtr{H_i H_1}I_{2;\what{i}}+\frac{1}{2} (D-3) \overtr{H_1 L}I_2\co\nn
     I_{\{1,2\}}&=-\frac{1}{4} (D-2) \overtr{L L}_{(i)} \overtr{H_i Z}\Big\vert_{z_2}I_{2;\what{i}}+\frac{1}{2} (D-3) \overtr{Z L}\Big\vert_{z_2}I_2\nn
     &=-\frac{1}{4} (D-2) \overtr{L L}_{(i)} \overtr{H_i H_2}I_{2;\what{i}}+\frac{1}{2} (D-3) \overtr{H_2 L}I_2\ed
     \end{align}
     Explicitly, using \eref{3-13}, we find the reduction coefficients are
     \begin{align}
     C_{\{2,1\}\to 2;\what{1}}&=\frac{D-2}{\left(\left(m_1-m_2\right){}^2-q_2^2\right) \left(\left(m_1+m_2\right){}^2-q_2^2\right)}\co\nn
     C_{\{2,1\}\to 2;\what{2}}&=-\frac{(D-2) \left(m_1^2+m_2^2-q_2^2\right)}{2 m_1^2 \left(\left(m_1-m_2\right){}^2-q_2^2\right) \left(\left(m_1+m_2\right){}^2-q_2^2\right)}\co\nn
     C_{\{2,1\}\to 2}&=\frac{(D-3) \left(m_1^2-m_2^2-q_2^2\right)}{\left(\left(m_1-m_2\right){}^2-q_2^2\right) \left(\left(m_1+m_2\right){}^2-q_2^2\right)}\co
     \end{align} 
     \begin{align}
     C_{\{1,2\}\to 2;\what{1}}&=-\frac{(D-2) \left(m_1^2+m_2^2-q_2^2\right)}{2 m_2^2 \left(\left(m_1-m_2\right){}^2-q_2^2\right) \left(\left(m_1+m_2\right){}^2-q_2^2\right)}\co\nn
     C_{\{1,2\}\to 2;\what{2}}&=\frac{D-2}{\left(\left(m_1-m_2\right){}^2-q_2^2\right) \left(\left(m_1+m_2\right){}^2-q_2^2\right)},\nn
     C_{\{1,2\}\to 2}&=-\frac{(D-3) \left(m_1^2-m_2^2+q_2^2\right)}{\left(\left(m_1-m_2\right){}^2-q_2^2\right) \left(\left(m_1+m_2\right){}^2-q_2^2\right)}\ed~~~\label{3-31}
     \end{align}
     \item \textbf{Tensor bubble with higher poles}
     
     At last, we consider a combined case, $I^{(1)}_{\mathbf{v}_2;v=3}$. Here we need to set $S=tV+Z,Z=z_1H_1+z_2H_2$. Setting $v=3,r=1,n=2$ in \eref{UniversalCoeff}, we have
     \begin{align}
     I^{(1)}_{\mathbf{v}_2;v=3}=-\Gamma \left(3-D/2\right)E_{2,4-D}[S^2]\Big\vert_{tz^{\mathbf{v}_2-1}}\ed
     \end{align}
     First, we use \eref{Enk-ViRecursion} iteratively to pull out all $S$'s in the numerator
     \begin{align}
     E_{2,4-D}[S^2]&=\a_{2,4-D}\left[\overtr{H_iS}E^{(i)}_{1,3-D}[S]+\overtr{SS}E_{2,2-D}+(2-D)\overtr{SL}E_{2,2-D}[S]\right]\nn
     &=\a_{2,4-D}\bigg[\b_{1,3-D}\overtr{H_iS}\overtr{SL}_{(i)}E^{(i)}_{1,1-D}+\overtr{SS}E_{2,2-D}\nn
     &\newline+(2-D)\overtr{SL}\a_{2,2-D}\big[\overtr{H_iS}E^{(i)}_{1,1-D}+(1-D)\overtr{SL}E_{2,-D}\big]\bigg]\ed
     \end{align}
     Among the four terms above, we only need to deal with the last term $E_{2,-D}$ since it corresponds to a $(D+2)$-dimensional bubble, which has been discussed in \eref{3-21}. Finally, we have
     \begin{align}
     I^{(1)}_{\mathbf{v}_2;v=3}&=-\frac{(D-2) \left(\overtr{L L} \overtr{H_i Z} \overtr{Z L}+\overtr{L L} \overtr{Z L}_{(i)} \overtr{H_i Z}-\overtr{H_i L} \overtr{Z L}^2\right)}{4 \overtr{L L}}\Big\vert_{tz^{\mathbf{v}_2-1}}I_{2;\what{i}}\nn
     &\newline +\frac{1}{2} \left(\frac{(D-2) \overtr{Z L}^2}{\overtr{L L}}-\overtr{Z Z}\right)\Big\vert_{tz^{\mathbf{v}_2-1}}I_2\ed
     \end{align}
     One can get  the reduction results for $I_{\{2,1\}}^{(1)},I_{\{1,2\}}^{(1)}$. For example,
     \begin{align}
     I^{(1)}_{\{2,1\}}&=-\frac{(D-2)}{4 \overtr{L L}} \bigg[\overtr{L L} \overtr{H_i V} \overtr{H_1 L}+\overtr{L L} \overtr{H_i H_1} \overtr{V L}+\overtr{L L} \overtr{V L}_{(i)} \overtr{H_i H_1}\nn
     &\newline +\overtr{L L} \overtr{H_1 L}_{(i)} \overtr{H_i V}-2\overtr{H_i L} \overtr{V L}\overtr{H_1 L}\bigg]I_{2;\what{i}}\nn
     &\newline + \left[\frac{(D-2) \overtr{H_1 L}\overtr{VL}}{\overtr{L L}}-\overtr{H_1 V}\right]I_2\co
     \end{align}

    where for simplicity, we will not present explicit expressions for these coefficients.
    \end{itemize}
    
    Note that the reduction coefficients in these examples are rational functions. For some special masses and momenta configurations,  denominators can become zero, which leads to several kinds of divergences. Since only $\overtr{LL}\equiv LQ^{-1}L$ appears in the denominators,  all divergences come from $Q$-matrix and its all sub-matrices, which have $\det Q=0$ or $\overtr{LL}=0$.  For example, the pole of $q_2^2$ in \eref{3-18}  comes from $LQ^{-1}L$ (see \eref{3-19}). The divergence of $C_{\{2,1\}\to 2;\what{2}}$ is given by $m_1^2 \left(\left(m_1-m_2\right){}^2-q_2^2\right) \left(\left(m_1+m_2\right){}^2-q_2^2\right)=0$, which is corresponds to $\det Q_{(2)}=0$ or $\det Q=0$.  One can find the pole $\overtr{LL}=0$ comes from the dimension shifting process \eref{Dimension-Lower}, which can be addressed  by employing \eref{Enk-V0Reduce} to reduce $E_{n,k}$ to lower topology. To deal with the divergences coming from $\det Q=0$, we need to consider the reduction method for degenerate $Q$ elaborated in the next section.

\section{Reduction for degenerate $Q$}
\label{sec:deQ-reduction}
In this section, we generalize our reduction method to degenerate $Q$. The basic idea is to generalize the recursion relation \eref{Base-Identity} to the formula 
\eref{A-9}. When $Q$ is degenerate, the characteristic equation $Q\xi=0$ always has solutions, and we denote the $\mathfrak{N}_Q$ as the null space spanned by linearly independent $\xi$'s.

\subsection{$\widetilde{Q}L\neq 0 $ }
When $Q$ is degenerate, the recursion relation \eref{Base-Identity} in the last sections breaks down for $\det{Q}=0$. Our idea is to consider the tensor structure with one $L$  in the first place
and make other $(k-1)$ indices completely symmetric by summing over all permutations between $i$ $V$'s and $(k-1-i)$ $L$'s. Using \eref{A-9} we have 
\bea
&&E_{n,k}[(Q\widetilde{Q}L)\otimes V^{i}\otimes L^{k-1-i}]\nn
&=&E_{n,k}\bigg[(Q\widetilde{Q}L)\otimes {1\over (k-1)!}\sum_{\sigma \in S_{k-1}}\sigma \left[V^i\otimes L^{k-1-i}\right]\bigg]\nn
&=&\a_{n,k}(H_b\widetilde{Q}L)E_{n-1,k-1}^{(b)}[V^i]+\b_{n,k}\left[{i\over k-1}(V\widetilde{Q}L)E_{n,k-2}[V^{i-1}]+{k-1-i\over k-1}(L\widetilde{Q}L)E_{n,k-2}[V^{i}]\right]\ed \label{SingQIdea}\nn
\eea

For a degenerate $Q$, we can always find a matrix $\widetilde{Q}=[\xi_1,\xi_2,\ldots,\xi_n]$,$\xi_i\in \mathfrak{N}_Q $ so that $Q\widetilde{Q}=0$. Then the LHS of \eref{SingQIdea} vanishes. If $Q^* L\neq 0$, where
$Q^*$ is the adjugate matrix of $Q$, we can take $\widetilde{Q}=Q^*$. With the denotation
\bea
\adovertr{AB}=(A\widetilde{Q}B)\co
\eea
\eref{SingQIdea} becomes
\bea
\a_{n,k}\adtr{H_bL}E_{n-1,k+1}^{(b)}[V^i]+i\b_{n,k}\adtr{VL}E_{n,k}[V^{i-1}]+(k+1-i)\b_{n,k}\adtr{LL}E_{n,k}[V^i]=0\ed \label{SingQBasic}
\eea
Depending on the value of $\adovertr{LL}$, we have following two cases:
\begin{itemize}
	\item (1) When $\adovertr{LL}\not=0$, \eref{SingQBasic} can be rewritten as
\bea
E_{n,k}^{\abs{Q}=0}[V^i]={1\over (i-k-1)\adovertr{LL}}\left[\adtr{H_bL}E_{n-1,k+1}^{(b)}[V^{i}]+i\adtr{VL}E_{n,k}[V^{i-1}]\right]\co \label{SingQnonLL}
\eea
where the first term in RHS corresponds to the lower topologies and the second term
has tensor rank reduced by one.

\item (2) When $\adovertr{LL}=0$, \eref{SingQBasic} can be rewritten as	
\bea
E_{n,k}^{\abs{Q}=0}[V^i]={-\adtr{H_bL}\over (i+1)\adtr{VL}}E^{(b)}_{n-1,k+1}[V^{i+1}]\co~~~\label{4-5}
\eea
where although the tensor rank increased by one on the RHS, it belongs to the lower
topologies. One particular thing of \eref{4-5} is that since $V$ depends on the
auxiliary $R$, we will always have $\adtr{VL}\neq 0$. Another tricky point is that although 
$R$ appears in the denominator in \eref{4-5} in the middle steps, it will be canceled
in the final reduction coefficients. 
\end{itemize}

\subsection{$\widetilde{Q}L=0$}
Since $\widetilde{Q}L=0$,  every term in \eref{SingQBasic} vanishes. Now we can put $V$ in the first place of tensor structure and using \eref{A-9} to reach
\bea
& & E_{n,k}[(Q\widetilde{Q}V)V^i]\nn &= & A_{n,k}\left [(H_j\widetilde{Q}V)E^{(j)}_{n-1,k-1}[V^{i}]+{i }(V\widetilde{Q}V)E_{n,k-2}[V^{i-1}]+(k-i-1)(V\widetilde{Q}L)E_{n,k-2}[V^i]\right]\ed
~~~\label{4-6}\eea
Using $Q\widetilde{Q}=\widetilde{Q}L=0$, \eref{4-6} becomes
\bea
0=A_{n,k}\left [(H_j\widetilde{Q}V)E^{(j)}_{n-1,k-1}[V^{i}]+{i }(V\widetilde{Q}V)E_{n,k-2}[V^{i-1}]\right]\co
\eea
and we have 
\bea
E_{n,k}[V^i]={-\adtr{H_bV}\over (i+1)\adtr{VV}}E_{n-1,k+1}^{(b)}[V^{i+1}]\ed \label{QtLzero}
\eea
Again, although $R$ appears in denominator through  $\adtr{VV}$ in middle steps,
 it will be canceled
in the final reduction coefficients. 
\subsection{Dimension recursion}
Having reduced to scalar integrals with different dimensions, we want to 
shift the dimension to a given $D$. Depending on various situations, we have: 

\begin{itemize}
	\item For $\widetilde{Q}L\neq 0$, we can always choose $Q\widetilde{Q}=0,\adtr{LL}\not =0 $. Then using  \eref{SingQnonLL} for the case 
	$i=0$,  the second term vanishes, and we have 

	\bea
	E_{n,k}={-\adtr{H_bL}\over (1+k)\adtr{LL}}E_{n-1,k+1}^{(b)}\ed \label{SingQnonLLDim}
	\eea

	\item For $\widetilde{Q}L=0$ we use \eref{QtLzero} with $i=1$. Then we
	 shift $V\to L+\eps K$ with a reference $K$ such that $K\not\perp \mathfrak{N}_Q$ and get
    \begin{align}
    E_{n,k}+\eps E_{n,k}[K]={-\adtr{H_bK}\over 2\eps\adtr{KK}}E_{n-1,k+1}^{(b)}[(L+\eps K)^2]\ed
    \end{align}
    Comparing both sides, especially the $\eps$ term, we have
    \begin{align}
    E_{n,k}={-\adtr{H_bK}\over \adtr{KK}}E_{n-1,k+1}^{(b)}[K]\ed ~~~\label{4-11}
    \end{align}
    We lower the topology in the RHS, so we can reduce it further by using the equation recursively.
    The dependence of the choice of $K$ in the middle steps will vanish in the final reduction coefficients as shown in later examples. 
    
\end{itemize}

\subsection{Examples}
In this section, we illustrate our method for degenerate $Q$. To avoid unnecessary and complicated calculations and compare with the reduction procedures for non-degenerate Q discussed in section \eref{sec:gQexamples}, we focus on  bubbles with some special masses and momenta configurations. 
\begin{itemize}
	\item \textbf{$\widetilde{Q}L= 0$: Massless scalar bubble with equal internal masses }
	
	To check the validity of our method, we first consider a scalar bubble with $m_1=m_2=m$ and $q_2^2=0$, defined as 
	\begin{align}
	I_{2}^{m_1=m_2,q_2^2=0}=\int {d^D\ell\over i \pi^{D/2}}{1\over (\ell^2-m^2)[(\ell-q_2)^2-m^2]}\co
	\end{align}
	which can be reduced to two tadpoles\footnote{One can check this by using FIRE or direct calculation. }.
	
	We find $Q$ defined in \eref{3-13} degenerates to a corank-1 matrix
	\begin{align}
	Q=\left[
	\begin{array}{cc}
	m^2 & m^2 \\
	m^2 & m^2 \\
	\end{array}
	\right]\co~~~\widetilde{Q}=Q^*=\left[
	\begin{array}{cc}
	m^2 & -m^2 \\
	- m^2 & m^2 \\
	\end{array}
	\right].
	\end{align}

	Since $Q\widetilde{Q}=0$ and $Q^*L=0$, 
	using \eref{E-I} we have
	\begin{align}
	I_{2}=\Gamma(2-D/2)E_{2,2-D}\ed\label{4-14}
	\end{align} 
 
 One can see the RHS is not irreducible anymore by employing \eref{4-11}
	\begin{align}
	E_{2,2-D}={-\adtr{H_bK}\over \adtr{KK}}E_{1,3-D}^{(b)}[K]\co\label{4-15}
	\end{align}
	where the reference vector $K$ satisfies $Q^*K\neq 0$  and the RHS corresponds to integrals of tadpole topology acquired by removing one propagator from $I_2$. We then use \eref{Enk-ViRecursion} to pull out the $K$
	\begin{align}
	E_{1,3-D}^{(b)}[K]=\beta_{1,3-D}\overtr{KL}_{(b)}E_{1,1-D}^{(b)}\ed\label{4-16}
	\end{align}
	Note that the $Q$-matrix in \eref{4-16}  is just a number $Q_{(b)}=m^2$, so its inverse in $\overtr{KL}_{(b)}$ is just $Q_{(b)}^{-1}={1\over m^2}$. Combining \eref{4-15},\eref{4-16} and \eref{4-14}, we finally find
	\begin{align}
	I_{2}^{m_1=m_2,q_2^2=0}={2-D\over 2}{\adtr{H_bK}\overtr{KL}_{(b)}\over \adtr{KK}}I_{2;\what{b}}={D-2\over 2m^2}I_{2;\what{1}}\ed
	\end{align}
	One can check that the result above does not depend on the choice of $K=(a,b)$ as long as $a\neq b$.

	\item \textbf{$\widetilde{Q}L= 0$: Massless tensor bubble with equal internal masses }
	
	Here we consider the reduction of the  tensor bubble  $I_{2}^{(1);m_1=m_2,q_2^2=0}$.  Recalling the first equation of \eref{Bubbler1Exp1}
    \begin{align}
    I_2^{(1)}=-{\Gamma(1-D/2)}(D-2) E_{2,2-D}[V]\co~~~\label{4-18}
    \end{align}
     we just need to  reduce $ E_{2,2-D}[V]$. Due to $\widetilde{Q}L= 0$ we use \eref{QtLzero} to get
    \begin{align}
    E_{2,2-D}[V]={-1\over 2\adtr{VV}}\adtr{H_bV}E_{1,3-D}^{(b)}[V^2]\co \label{deQ-bub-r1}
    \end{align}
    where the RHS corresponds to integrals of tadpoles  with  non-degenerate $Q$-matrix $Q_{(b)}=m^2$. We can use \eref{Enk-ViRecursion} reduce it iteratively
    \begin{align}
    E_{1,3-D}^{(b)}[V^2]&=\a_{1,3-D}\left[\overtr{VV}_{(b)}E^{(b)}_{1,1-D}+(1-D)\overtr{VL}_{(b)}E^{(b)}_{1,1-D}[V]\right]\nn
    &=\a_{1,3-D}\left[\overtr{VV}_{(b)}E^{(b)}_{1,1-D}+(1-D)\overtr{VL}_{(b)}\beta_{1,1-D}\overtr{VL}_{(b)}E_{1,-1-D}^{(b)}\right]\ed\label{deQ-tad-r1}
    \end{align}
    There are two terms in the last line, we need to reduce the second one $E_{1,-1-D}^{(b)}$ since it corresponds to $(D+2)$-dimensional tadpole. Using \eref{Dimension-Lower} to lower the dimension to $D$, we get
    \begin{align}
    E^{(b)}_{1,-1-D}={E^{(b)}_{1,1-D}\over \b_{1,1-D}\overtr{LL}_{(b)}}\ed \label{tad-dim-higher}
    \end{align}
    Plugging \eref{deQ-tad-r1},\eref{tad-dim-higher}, \eref{deQ-bub-r1} into \eref{4-18}, we have
    \begin{align}
    I_{2}^{(1);m_1=m_2,q_2^2=0}&={\adtr{H_bV}\left((1-D)\overtr{VL}_{(b)}^2+\overtr{LL}_{(b)}\overtr{VV}_{(b)}\right)\over 2\adtr{VV} \overtr{LL}_{(b)}}I_{2;\what{b}}\ed
    \end{align}
    Using
    \begin{align}
    \adtr{VV}&=m^2 (R\cdot q_2)^2\co~\hspace{22pt}\adtr{H_bV}=\left\{-m^2R\cdot q_2,m^2R\cdot q_2\right\}\co \nn
    \overtr{VV}_{(b)}&=\left\{{(R\cdot q_2)^2\over m^2},0\right\}\co~~~\overtr{VL}_{(b)}=\left\{{R\cdot q_2\over m^2},0\right\}\co ~~~\overtr{LL}_{(b)}={1\over m^2}\co
    \end{align}
    we find the  reduction relation
    \begin{align}
    I_{2}^{(1);m_1=m_2,q_2^2=0}={(D-2)R\cdot q_2\over 2m^2}I_{2;\what{1}}\co
    \end{align}
    where $I_{2;\what{1}}$ is just the tadpole $I_1[m]$ with mass $m$.

    \item \textbf{$\widetilde{Q}L\neq  0$: Scalar bubble }
    
    We then consider the scalar bubble $I_2^{(1)}$ with degenerate $Q$ but $m_1\not =m_2$. 
    Here we will show that it can be reduced to two tadpoles using our method. The equation $\det Q=0$ gives two solutions
    \bea
    q_2^2=(m_1\pm m_2)^2\co~~~\label{4-25}
    \eea
    which are just the poles in the reduction coefficients of bubbles (see \eref{3-31})\footnote{The condition is the threshold (or the Landau poles) of the integrals. It appears is because with higher power of propagators, the singularity of Landau poles becomes higher. Thus, it will appear in both master integrals and the reduction coefficients.}. Here we choose
    \bea
    \widetilde{Q}=Q^*=\left(
    \begin{array}{cc}
    	m_2^2 &\pm m_1 m_2 \\
    	\pm 	m_1 m_2 & m_1^2 \\
    \end{array}
    \right)\ed
    \eea
    One can check that $\widetilde{Q}L\not =0$ for $m_1\not =m_2$.
    Due to $\adtr{LL}\not=0$, we can use \eref{SingQnonLL} to reduce $E_{2,2-D}$ in the expansion
    \begin{align}
    I_{2}=\Gamma(2-D/2)E_{2,2-D}\co\label{4-28}
    \end{align} 
    and we get
    \begin{align}
    E_{2,2-D}={-\adtr{H_bL}\over (3-D)\adtr{LL}}E_{1,3-D}^{(b)}={\adtr{H_bL}\b_{1,3-D}\over (D-3)\adtr{LL}}\overtr{LL}_{(b)}E_{1,1-D}^{(b)}\ed\label{4-29}
    \end{align}
    Plugging \eref{4-29} into \eref{4-28}, we have
    \begin{align}
    I_{2}^{m_1=m_2,q_2^2=0}={(D-2)\adtr{H_bL}\overtr{LL}_{(b)}\over 2(D-3)\adtr{LL}}I_{2;\what{b}}\ed\label{4-30}
    \end{align}
    One can find the explicit expression easily.  For $q_2^2=(m_1+m_2)^2$, the explicit expression is 
    \begin{align}
    I_{2}^{m_1=m_2,q_2^2=0}=\frac{D-2}{2 (D-3) m_2 \left(m_1+m_2\right)}I_{2;\what{1}}+\frac{D-2}{2 (D-3) m_1 \left(m_1+m_2\right)}I_{2;\what{2}}\ed\label{4-31}
    \end{align}

    \item \textbf{$\widetilde{Q}L\neq  0$: Tensor bubble }
    
    Here we discuss the reduction of the rank-1 bubble with $q_2^2=(m_1\pm m_2)^2$. First, we expand it using \eref{UniversalCoeff}
    \begin{align}
    I_2^{(1)}=-{\Gamma(1-D/2)}(D-2) E_{2,2-D}[V]\ed\label{4-32}
    \end{align}
    Choosing $\widetilde{Q}=Q^*$, due to $\adtr{LL}=LQ^*L\not =0$, we can use \eref{SingQnonLL} to reduce the RHS to
    \begin{align}
    E_{2,2-D}[V]={1\over (D-2)\adovertr{LL}}\left[\adtr{H_bL}E_{1,3-D}^{(b)}[V]+\adtr{VL}E_{2,2-D}\right]\co\label{4-33}
    \end{align}
    where the second term has been discussed in the last example. We just need to reduce the first term $E_{1,3-D}^{(b)}[V]$, which corresponds to a tadpole. Using \eref{Enk-ViRecursion}, we find
    \begin{align}
    E_{1,3-D}^{(b)}[V]=\b_{1,3-D}\overtr{VL}_{(b)}E_{1,1-D}^{(b)}\co\label{4-34}
    \end{align}
    where the RHS is a $D$-dimensional scalar tadpole. Plugging \eref{4-34},\eref{4-33} into \eref{4-32} we get
    \begin{align}
    I_2^{(1);q_2^2=(m_1\pm m_2)^2}={\adtr{H_bL}\overtr{VL}_{(b)}\over \adtr{LL}}I_{2;\what{b}}+{2\over D-2}{\adtr{VL}\over \adtr{LL}}I_2^{q_2^2=(m_1\pm m_2)^2}\ed
    \end{align}
    Then we refer to the result \eref{4-30}, and find
    \begin{align}
    I_2^{(1);q_2^2=(m_1\pm m_2)^2}={\adtr{H_bL}\overtr{VL}_{(b)}\over \adtr{LL}}I_{2;\what{b}}+{\adtr{H_bL}\overtr{LL}_{(b)}\adtr{VL}\over (D-3)\adtr{LL}^2}I_{2;\what{b}}\ed
    \end{align}
   For $q_2^2=(m_1+m_2)^2$, the explicit expression is 
   \bea
   I_2^{(1);q_2^2=(m_1+ m_2)^2}=\frac{\left((D-2) m_1+(D-3) m_2\right) R\cdot q_2}{(D-3) m_2 \left(m_1+m_2\right){}^2}I_{2;\what{1}}+\frac{ R\cdot q_2}{(D-3)  \left(m_1+m_2\right){}^2}I_{2;\what{2}}\ed
   \eea
\end{itemize}

\section{General expression of $C^{(r)}_{n\to n}$}
\label{sec:ngonself}
It seems hard to solve \eref{Enk-ViRecursion}, but if we only care the reduction coefficients to the same topology, we can ignore the first term for it  contributes only to lower topologies. Then by iteratively using 
\eref{Enk-ViRecursion}, keeping only the second and the third term, one find that 

\bea
E_{n,k}[V^{i}]=\sum_{j}{i!\over j!(i-j)!!}\overtr{VV}^{{i-j\over 2}}\overtr{VL}^{j}{\prod_{l=1}^{j}(k-i-l+1)\over \prod_{l=1}^{{i+j\over 2}}(k+n-2l)}E_{n,k-i-j}+\textrm{Lower topology}\ed~~~\label{5-1}
\eea
Although the first term has the same topology, with different choices of $i,j$, the dimension is different, thus we need to reduce it further. 

To simplify our denotation, we define
\bea
\mathscr{E}_{n,k,i;j}={i!\over j!(i-j)!!}{(k-i)!(k+n-i-j-2)!!\over (k-i-j)!(k+n-2)!!}\ed
\eea
Thus \eref{5-1} becomes
\bea
E_{n,k}[V^i]=\sum_j\mathscr{E}_{n,k,i;j}\overtr{VV}^{{i-j\over 2}}\overtr{VL}^jE_{n,k-i-j}+\textrm{Lower topology}\ed
\eea
Here we give the computation for tensor reduction with all propagators power one\footnote{The general results for arbitrary tensor structure,  general powers as well as the coefficients for lower topologies are given in  Appendix \ref{apdix:generalcoeff}.}. Thus, we need to reduce all $E_{n,n-D-(r-i)}[V^i]$ to $E_{n,n-D}$. After using \eref{Dimension-Lower} to repeatedly lift $k$ we have
\bea
E_{n,k-2s}&=&{E_{n,k}\over \overtr{LL}^{s}\prod_{l=1}^{s} (\b_{n,k-2s+2l})}+\sum_{l=1}^s{-\a_{n,k-2s+2l}\overtr{H_jL}\over \prod_{p=1}^{l}\b_{n,k-2s+2p}\overtr{LL}^l}E^{(j)}_{n-1,k-1-2s+2l}\nn
&=&{\mathscr{K}^+_{n,k,s}\over \overtr{LL}^s}E_{n,k}+\sum_{j=1}^s{\mathscr{K}^+_{n,k,s;j}\over \overtr{LL}^j}\overtr{H_bL}E^{(b)}_{n-1,k-1-2(s-j)}\co~~~~\label{5-4}
\eea
where we have defined 
\bea
\mathscr{K}^+_{n,k,s}={1\over \prod_{l=1}^{s} (\b_{n,k-2s+2l})}\co~~~~~\mathscr{K}^+_{n,k,s;j}={-\a_{n,k-2s+2j}\over \prod_{p=1}^{j}\b_{n,k-2s+2p}}\ed
\eea
Since the second term in \eref{5-4} belongs to lower topologies, which can be ignored,  finally we have
\bea
& & E_{n,n-D-(r-i)}[V^{i}]=\sum_j\mathscr{E}_{n,n-D-(r-i),i;j}\overtr{VV}^{i-j\over 2}\overtr{VL}^jE_{n,n-D-(r+j)}+Lower\nn
&=&\sum_j\mathscr{E}_{n,n-D-(r-i),i;j}\overtr{VV}^{i-j\over 2}\overtr{VL}^j{\mathscr{K}^+_{n,n-D,{r+j\over 2}}\over \overtr{LL}^{r+j\over 2}}E_{n,n-D}+Lower\nn
&=&\sum_j\mathscr{E}_{n,n-D-(r-i),i;j}\overtr{VV}^{i-j\over 2}\overtr{VL}^j{\mathscr{K}^+_{n,n-D,{r+j\over 2}}\over \overtr{LL}^{r+j\over 2}}{(-)^n\over \Gamma(n-D/2)}I_{n;D}+Lower\ed
\eea
So
\bea
I_{n;D}^{(r)}&=&\sum_{i=0}^{r} {\Gamma(n-D/2-r)\over (-1)^{n+r}  }\mathscr{C}^{D/2+r-n}_{r,i}(R^2)^{r-i\over 2} E_{n,n-D-(r-i)}[V^i]\nn
&=&\sum_{i=0}^{r} {\Gamma(n-D/2-r)\over (-1)^{n+r}  }\mathscr{C}^{D/2+r-n}_{r,i}(R^2)^{r-i\over 2}\times \nn
&& \sum_j\mathscr{E}_{n,n-D-(r-i),i;j}\overtr{VV}^{i-j\over 2}\overtr{VL}^j{\mathscr{K}^+_{n,n-D,{r+j\over 2}}\over \overtr{LL}^{r+j\over 2}}{(-)^n\over \Gamma(n-D/2)}I_{n;D}+Lower\nn
&=&{(-)^r\Gamma(n-D/2-r)\over \Gamma(n-D/2)}\sum_{i=0}^r\sum_{j=0}^i\mathscr{C}^{D/2+r-n}_{r,i}\mathscr{E}_{n,n-D-(r-i),i;j}\mathscr{K}^+_{n,n-D,{r+j\over 2}}\times \nn
&&(R^2)^{r-i\over 2}{\overtr{VV}^{i-j\over 2}\overtr{VL}^j\over \overtr{LL}^{r+j\over 2}}I_{n;D}+Lower\ed
\eea
From it, we read out reduction coefficient 
\bea
C^{(r)}_{n\to n}&=&\sum_{i=0}^r\sum_{j=0}^ic^{(r)}_{n\to n;i,j}(R^2)^{r-i\over 2}{\overtr{VV}^{i-j\over 2}\overtr{VL}^j\over \overtr{LL}^{r+j\over 2}}\label{Coeff-n-n}
\eea
with 
\bea
c^{(n)}_{n\to n;i,j}={(-)^r\Gamma(n-D/2-r)\over \Gamma(n-D/2)}\mathscr{C}^{D/2+r-n}_{r,i}\mathscr{E}_{n,n-D-(r-i),i;j}\mathscr{K}^+_{n,n-D,{r+j\over 2}}\co
\eea
where we require $r,i,j$ having the same parity.

\section{Discussion}
Although the main target of the paper~\cite{Arkani-Hamed:2017ahv} is to understand general one-loop integrals from a geometric point of view, it does contain many important and valuable results. In this paper, we have elaborated on the method to compute reduction coefficients for general one-loop integrals. 

The essential idea of the method is to put the whole one-loop Feynman integrals in the projective space, in which integrals have compact forms and geometry properties. Using a vital recursion relation of $E_{n,k}[T]$, one achieves the wanted reduction. The advantage and the most promising point of this method is that we can solve any one-loop integrals with higher poles and tensor structures at the same time, which is demonstrated by some examples in section \ref{sec:one-loop reduction} and \ref{sec:deQ-reduction}. The language of projective space simplifies the reduction process a lot by keeping the elegant contractions like $\overtr{HL},\overtr{LL}$ in these recursion relations without expansion, thus making the whole reduction process a symbolic calculation. For some programs based on the  traditional IBP method, like {\sc FIRE}, {\sc LiteRed},{\sc Kira}, etc \cite{Smirnov:2008iw,Smirnov:2014hma,Lee:2013mka,Smirnov:2019qkx,vonManteuffel:2012np,Maierhofer:2017gsa,Gerlach:2022qnc}, they do reduction by solving linear equations where the determinant of the Gram matrix appears with the full-expanded form, which makes the final results too complicated to read. The appearance of (reduced) Gram matrix
has also been observed in our recent work \cite{Feng:2021enk,Hu:2021nia}.

One obvious idea is to generalize the above method to the reduction of two-loops and higher-loops. Recently, using the improved PV-reduction method with auxiliary vectors, we have shown how to do the general tensor reduction for two-loop sunset integrals \cite{Feng:2022iuc}. From the results in \cite{Feng:2021enk,Hu:2021nia} and results in this paper, we see that these two methods have some correspondences. In other words, they treat the same thing from different but related angles. Since the improved PV-reduction method has some results for two-loop reductions, the projective space method should be generalized to two-loop too.

\section*{Acknowledgments}
This work is supported by Qiu-Shi Funding and Chinese NSF funding under Grant No.11935013, No.11947301, No.12047502 (Peng Huanwu Center).

\appendix

\section{Recursion relations of $E_{n,k}[T]$}
\label{sec:proof of Enk}
In this section, we recall the proof of the relation \eref{Base-Identity} given in \cite{Arkani-Hamed:2017ahv}. 
Let us calculate directly
\bea
&&\dd_X \left[{\langle Q^{-1}T[X^{k-1}]X\dd^{n-2}X\rangle \over (XQX)^{{n+k-2\over 2}}}\right]= {1\over (XQX)^{{n+k-2\over 2}}}\times \nn
&&\Big[-(n+k-2){(XQ\dd X)\over (XQX)}\langle Q^{-1}T[X^{k-1}]X\dd^{n-2}X\rangle +(n-1)\langle Q^{-1}T[X^{k-1}]\dd^{n-1} X\rangle\nn
&+&(k-1)X_{I_1}{1\over (n-2)!}Q^{-1}_{I_1i_1}T^{i_1i_2i_3\ldots i_k}X_{i_2}X_{i_3}\cdots \dd X_{i_k}X_{I_2}\wedge \dd^{n-2}X \eps^{I_1I_2\ldots I_k}\Big]\co~~~\label{42-proof-1}
\eea
where we require $T$ is completely symmetric of the last $(k-1)$ indices $i_2i_3\ldots i_k$. For simplicity, we denote
\bea
A_I=Q^{-1}_{Is} T^{s i_2i_3\ldots i_k}X_{i_2}X_{i_3}\cdots X_{i_k}\co~~~~ B_{Ir}=(Q^{-1}T[X^{k-2}])_{Ir}\ed~~~\label{42-proof-2}
\eea
Thus \eref{42-proof-1} becomes
\bea
&&\dd_X \left[{\langle Q^{-1}T[X^{k-1}]X\dd^{n-2}X\rangle \over (XQX)^{{n+k-2\over 2}}}\right]=  {1\over (XQX)^{{n+k-2\over 2}}}\times \nn & & 
\Big[-(n+k-2){(XQ\dd X)\over (XQX)}\langle AX\dd^{n-2}X\rangle +(n-1)\langle A\dd^{n-1} X\rangle+(k-1)\dd X_I\langle B^{\ I}X\dd^{n-2}X\rangle \Big]\ed
~~~~~~\label{42-proof-3}\eea
Using the fact 
\bea
XQA=XQQ^{-1}T[X^{k-1}]=T[X^k]\co~~~~B_I^{\ I}=(tr_QT)[X^{k-2}]\co
\eea
and 
the Schouten identity 
\begin{align}
\dd X^I \langle AX\dd X^{n-2}\rangle+A^I\langle X\dd^{n-1}X\rangle-X^I\langle A\dd^{n-1}X\rangle =0\co~~~~\forall A^I
\end{align}
to simplify the first term in \eref{42-proof-3} as 
\bea
&&XQ\dd X \langle AX\dd^{n-2}X \rangle=QX\dd X\left(-A\langle X\dd^{n-1}X\rangle +X \langle A\dd^{n-1}X\rangle \right)\nn
&&=-(XQA)\langle X\dd^{n-1}X\rangle +(QXX)\langle A \dd^{n-1}X \rangle\co
\eea
and the third term in \eref{42-proof-3} as 
\bea
\dd X_I\langle B^{\ I}X\dd^{n-2}X\rangle= X^I \langle B_{\ I}\dd^{n-1}X \rangle -B_I^{\ I}\langle X \dd^{n-1}X \rangle\co
\eea
the  \eref{42-proof-3} becomes
\bea
\dd_X \left[{\langle Q^{-1}T[X^{k-1}]X\dd^{n-2}X\rangle \over (XQX)^{{n+k-2\over 2}}}\right]=(n+k-2){T[X^k]\langle X\dd^{n-1}X \rangle \over (XQX)^{{n+k\over 2}}}-(k-1)tr_QT[X^{k-2}]\measure{X}{n-1}\co~~~~~~\label{42-proof-4}
\eea 
which is nothing but the wanted \eref{Base-Identity}. 

The derivation above can be generalized to the case ${\rm det}Q=0$, where the $Q^{-1}$ does not exist. In this
case, we can replace $Q^{-1}$ by arbitrary matrix $\widetilde{Q}$ in \eref{42-proof-1} and repeat the same derivations to reach the similar expression likes  \eref{42-proof-4}, except $Q^{-1}$ replaced by $\widetilde{Q}$. Rearranging \eref{42-proof-4} we get 
\bea
E_{n,k}[(Q\widetilde{Q}T)]=\alpha_{n,k}E^{(b)}_{n-1,k-1}[(H_b\widetilde{Q}T)]+\beta_{n,k}E_{n,k-2}[tr_{\widetilde{Q}} T]\co~~~~\label{A-9}
\eea
where $(Q\widetilde{Q}T)$ has the same rank as $T$
\begin{align}
(Q\widetilde{Q}T)^{I_1I_2,\ldots,I_k}=Q^{I_1J_1}\widetilde{Q}_{J_1J_2}T^{J_2,I_2,I_3,\ldots,I_k}\co~~~~(tr_{\widetilde{Q}} T)^{I_2,I_3,\ldots,I_k}=\widetilde{Q}_{I_1I_2}T^{I_1I_2\ldots I_k}\ed
\end{align}
In \eref{A-9}, the LHS is the first term in RHS of \eref{42-proof-4},
while the first term on the RHS is the boundary contribution of LHS of \eref{42-proof-4}.
\section{More results}
\label{sec:more results}

\subsection{Bubbles}
Here list the results for rank from $r=1$ to $r=4$, where we set $q_1=0$.
\begin{itemize}
	\item $r=1$
	\bea
	I_2^{(1)}=\sum_i-\frac{\overtr{L L} \overtr{H_i V}-\overtr{H_i L} \overtr{V L}}{\overtr{L L}}I_{{2;\widehat{i}}}
	+ \frac{2 \overtr{V L}}{\overtr{L L}}I_2\ed
	\eea
	So we have
	\bea
	C^{(1)}_{2\to 2;\widehat{2}}&=&-\frac{\overtr{L L} \overtr{H_2 V}-\overtr{H_2 L} \overtr{V L}}{\overtr{L L}}=-\frac{R\cdot q_2}{q_2^2}\co\nn
	C^{(1)}_{2\to 2;\widehat{1}}&=&-\frac{\overtr{L L} \overtr{H_1 V}-\overtr{H_1 L} \overtr{V L}}{\overtr{L L}}=\frac{R\cdot q_2}{q_2^2}\co\nn
	C^{(1)}_{2\to 2}
	&=&\frac{2 \left(VQ^{-1}L\right)}{\left(LQ^{-1}L\right)}=\frac{\left(m_1^2-m_2^2+q_2^2\right) R\cdot q_2}{q_2^2}\ed~~~~\label{Bub-r=1-coeff}
	\eea
	\item $r=2$\\
	\bea
	C^{(2)}_{2\to 2;{\widehat{i}}}&=&\frac{2 R^2 \overtr{H_i L}}{(D-1) \overtr{L L}}+\frac{2 D \overtr{H_i L} \overtr{V L}^2}{(D-1) \overtr{L L}^2}-\frac{2 \overtr{H_i V} \left(\overtr{V L}+\overtr{V L}_{(i)}\right)}{\overtr{L L}_{(i)}}\nn
	&&-\frac{2 \overtr{H_i L} \big(-D \overtr{V L}^2+\overtr{V L}^2+\overtr{L L}_{(i)} \overtr{V V}\big)}{(D-1) \overtr{L L}_{(i)} \overtr{L L}}\co\nn
	C^{(2)}_{2\to 2}
	&=&\frac{4 \left(R^2-\overtr{V V}\right)}{(D-1) \overtr{L L}}+\frac{4 D \overtr{V L}^2}{(D-1) \overtr{L L}^2}\ed~~~~\label{App-B-bub-r=2}
	\eea
	the exact form is 
	\bea
	C^{(2)}_{2\to 2;\widehat{2}}&=&\frac{\left(m_1^2-m_2^2+q_2^2\right) R^2}{(D-1) q_2^2}-\frac{D \left(m_1^2-m_2^2+q_2^2\right) \left(R\cdot q_2\right){}^2}{(D-1) q_2^4}\co\nn
	C^{(2)}_{2\to 2;\widehat{1}}&=&-\frac{R^2 \left(m_1^2-m_2^2-q_2^2\right)}{(D-1) q_2^2}-\frac{\left(-D m_1^2+D m_2^2-3 D q_2^2+4 q_2^2\right) \left(R\cdot q_2\right){}^2}{(D-1) q_2^4}\co\nn
	C^{(2)}_{2\to 2}&=&-\frac{R^2 \left(-2 m_1^2 \left(m_2^2+q_2^2\right)+\left(m_2^2-q_2^2\right){}^2+m_1^4\right)}{(D-1) q_2^2}\nn
	&&+\frac{\left(-2 m_1^2 \left(D m_2^2-(D-2) q_2^2\right)+D \left(m_2^2-q_2^2\right){}^2+D m_1^4\right) \left(R\cdot q_2\right){}^2}{(D-1) q_2^4}\ed\no
	\eea

	\item $r=3$\\

	\begin{align}
	C^{(3)}_{2\to 2;{\widehat{i}}}&=
	-\frac{4 (D+1) \overtr{H_i V} \left(\overtr{V L}_{(i)} \overtr{V L}+\overtr{V L}^2+\overtr{V L}_{(i)}^2\right)}{D \overtr{L L}_{(i)}^2}\nn
	&\newline-\frac{4 \overtr{H_i V} \left(-2 \overtr{V V}-\overtr{V V}_{(i)}+3 R^2\right)}{D \overtr{L L}_{(i)}}+\frac{4 (D+2) \overtr{H_i L} \overtr{V L}^3}{(D-1) \overtr{L L}^3}\nn
	&\newline+\frac{4 \overtr{H_i L} \overtr{V L} \left(D \overtr{V L}^2+\overtr{V L}^2-3 \overtr{L L}_{(i)} \overtr{V V}+3 R^2 \overtr{L L}_{(i)}\right)}{D \overtr{L L}_{(i)}^2 \overtr{L L}}\nn
	&\newline +\frac{4 \overtr{H_i L} \overtr{V L} }{(D-1) D \overtr{L L}_{(i)} \overtr{L L}^2}\Big(D^2 \overtr{V L}^2+D \overtr{V L}^2-3 D \overtr{L L}_{(i)} \overtr{V V}\nn
	&\newline \newline-2 \overtr{V L}^2+3 D R^2 \overtr{L L}_{(i)}\Big)\co\nn
	C^{(3)}_{2\to 2}
	&=\frac{24 R^2 \overtr{V L}}{(D-1) \overtr{L L}^2}+\frac{8 (D+2) \overtr{V L}^3}{(D-1) \overtr{L L}^3}-\frac{24 \overtr{V V} \overtr{V L}}{(D-1) \overtr{L L}^2}\ed
	\end{align}

	The exact form is 
	\allowdisplaybreaks
	\begin{align}
	C^{(3)}_{2\to 2;\widehat{2}}&=-\frac{(D+2) \left(m_1^2-m_2^2\right){}^2 \left(R\cdot q_2\right){}^3}{(D-1) q_2^6}+\frac{3 R^2 R\cdot q_2}{D-1}\nn
	&\newline+\frac{\left(2 D (D+2) m_2^2-2 \left(D^2-2 D+4\right) m_1^2\right) \left(R\cdot q_2\right){}^3+3 D \left(m_1^2-m_2^2\right){}^2 R^2 R\cdot q_2}{(D-1) D q_2^4}\nn
	&\newline-\frac{R\cdot q_2 \left(6 R^2 \left((D-2) m_1^2+D m_2^2\right)+D (D+2) \left(R\cdot q_2\right){}^2\right)}{(D-1) D q_2^2}\co\nn
	C^{(3)}_{2\to 2;\widehat{1}}&=
	\frac{3 R^2 \left(4 (D-1) m_2^2 q_2^2-D m_1^4+2 D m_2^2 m_1^2-D m_2^4+D q_2^4\right) R\cdot q_2}{(D-1) D q_2^4}\nn
	&\newline\frac{\left(R\cdot q_2\right){}^3}{(D-1) D q_2^6}\Big[-4 \left(D^2+D-2\right) m_2^2 q_2^2-2 D m_1^2 \left((D+2) m_2^2-2 (D-1) q_2^2\right)\nn
	&\newline+D (D+2) m_1^4+D (D+2) m_2^4+D (7 D-10) q_2^4\Big]\co\nn
	C^{(3)}_{2\to 2}&=\frac{\left(m_1^2-m_2^2+q_2^2\right)  \left(R\cdot q_2\right){}^3}{(D-1) q_2^6}\Big [-2 m_1^2 \left((D+2) m_2^2-(D-4) q_2^2\right)\nn
	&\newline+(D+2) \left(m_2^2-q_2^2\right){}^2+(D+2) m_1^4\Big]\nn
	&\newline-\frac{3 R^2 \left(m_1^2-m_2^2+q_2^2\right) \left(-2 m_1^2 \left(m_2^2+q_2^2\right)+\left(m_2^2-q_2^2\right){}^2+m_1^4\right) R\cdot q_2}{(D-1) q_2^4}\ed
	\end{align}
	\item $r=4$\\
	\allowdisplaybreaks
	\begin{align}
	C^{(4)}_{2\to 2;\widehat{i}}&=\frac{8 (D+2) \overtr{H_i L} \overtr{V L}^2 \left(\left(D^2+3 D-4\right) \overtr{V L}^2+6 D \overtr{L L}_{(i)} \left(R^2-\overtr{V V}\right)\right)}{(D-1) D (D+1) \overtr{L L}_{(i)} \overtr{L L}^3}\nn
	&\newline+\frac{8 \overtr{H_i L} }{D (D+1) \overtr{L L}_{(i)}^3 \overtr{L L}}\Big(\left(D^2+4 D+3\right) \overtr{V L}^4+3 \overtr{L L}_{(i)}^2 \left(R^2-\overtr{V V}\right){}^2\nn
	&\newline+6 (D+1) \overtr{L L}_{(i)} \overtr{V L}^2 \left(R^2-\overtr{V V}\right)\Big)+\frac{8 \overtr{H_i L} }{D \left(D^2-1\right) \overtr{L L}_{(i)}^2 \overtr{L L}^2}\times \nn
	&\newline \Big(\left(D^3+4 D^2-D-4\right) \overtr{V L}^4 +6 \left(D^2+D-2\right)\overtr{L L}_{(i)} \overtr{V L}^2 \left(R^2-\overtr{V V}\right)\nn&\newline+3 D \overtr{L L}_{(i)}^2 \left(R^2-\overtr{V V}\right){}^2\Big)+\frac{8 (D+2) (D+4) \overtr{H_i L} \overtr{V L}^4}{\left(D^2-1\right) \overtr{L L}^4}\nn
	&\newline-\frac{8 \overtr{H_i V} }{D \overtr{L L}_{(i)}^2}\Big(\overtr{V L} \left(-5 \overtr{V V}-\overtr{V V}_{(i)}+6 R^2\right)+\overtr{V L}_{(i)} \left(-3 \overtr{V V}-3 \overtr{V V}_{(i)}+6 R^2\right)\Big)\nn
	&\newline-\frac{8 (D+3)\left(\overtr{V L}+\overtr{V L}_{(i)}\right)\left(\overtr{V L}^2+\overtr{V L}_{(i)}^2\right)  \overtr{H_i V} }{D \overtr{L L}_{(i)}^3}\co\nn
	C^{(4)}_{2\to 2}&=\frac{96 (D+2) \overtr{V L}^2 \left(R^2-\overtr{V V}\right)}{\left(D^2-1\right) \overtr{L L}^3}+\frac{48 \left(R^2-\overtr{V V}\right){}^2}{\left(D^2-1\right) \overtr{L L}^2}+\frac{16 \left(D^2+6 D+8\right) \overtr{V L}^4}{\left(D^2-1\right) \overtr{L L}^4}\ed
	\end{align}

\end{itemize}
\subsection{Triangles}
For triangle topology, we have presented results for scalar triangles with higher poles. Here we present some examples, including the tensor triangles without higher poles and with higher poles. 

For the tensor triangles without higher poles, we have:
\begin{itemize}
	\item $r=1$:
	\bea
	\tensorc{1}{3}{i}&=&\frac{\overtr{H_i L} \overtr{V L}}{\overtr{L L}}-\overtr{H_i V}\co\nn
	\tensorselfc{1}{3}&=&\frac{2 \overtr{V L}}{\overtr{L L}}\ed
	~~~~\label{App-B-tri-1-1}
	\eea
	The expression is the same as \eref{Bub-r=1-coeff}
	for bubble topology. This phenomenon is not an accident,
	and will be persistent to other topologies. 
	\item $r=2$:
	\bea
	\tensorc{2}{3}{ij}&=&\frac{\overtr{H_i L} \overtr{H_j L}_{(i)} \overtr{V L}^2}{\overtr{L L}_{(i)} \overtr{L L}}-\frac{\overtr{H_j L}_{(i)} \overtr{H_i V} \left(\overtr{V L}_{(i)}+\overtr{V L}\right)}{\overtr{L L}_{(i)}}\nn
	&&+\overtr{H_i V} \overtr{H_j V}_{(i)}+(i\leftrightarrow j)\co\nn
	\tensorc{2}{3}{i}&=&\frac{2 \overtr{H_i L} \left((D-2) \overtr{V L}^2+\overtr{L L}_{(i)} \left(R^2-\overtr{V V}\right)\right)}{(D-2) \overtr{L L}_{(i)} \overtr{L L}}\nn
	&&+\frac{2 (D-1) \overtr{H_i L} \overtr{V L}^2}{(D-2) \overtr{L L}^2}\-\frac{2 \overtr{H_i V} \left(\overtr{V L}_{(i)}+\overtr{V L}\right)}{\overtr{L L}_{(i)}}\co\nn
	\tensorselfc{2}{3}&=&\frac{4 \left((D-1) \overtr{V L}^2+\overtr{L L} \left(R^2-\overtr{V V}\right)\right)}{(D-2) \overtr{L L}^2}\ed~~~~\label{App-B-tri-1-2}
	\eea
	Again, the last two coefficients are very similar to
	these given in \eref{App-B-bub-r=2}. 
\end{itemize}

For the tensor with higher poles, we have: 
\begin{itemize}
	\item $v=4,r=1$:
	\allowdisplaybreaks
	\bea
	C^{(1)}_{\mathbf{v}_3 \to 3;\widehat{ij}}&=&\frac{1}{8} (D-2) \overtr{L L}_{(ij)} \left(\overtr{H_i S} \overtr{H_j S}_{(i)}+(i \leftrightarrow j)\right)\co\nn
	C^{(1)}_{\mathbf{v}_3 \to 3;\widehat{i}}&=&{D-3\over 4} \left(\frac{\overtr{H_i L} \overtr{S L}^2}{\overtr{L L}}- \overtr{H_i S} \overtr{S L}_{(i)}-\overtr{H_i S} \overtr{S L}\right)\co\nn
	C^{(1)}_{\mathbf{v}_3 \to 3}&=&\frac{(D-3) \overtr{S L}^2}{2 \overtr{L L}}-\frac{1}{2} \overtr{S S}\ed~~~~\label{App-B-tri-2-1}
	\eea
	Here we have suppressed the process to take the coefficient of $tz^{\mathbf{v}_3-1}$,i.e, $\vert_{tz^{\mathbf{v}_3-1}}$. The similarity of the above expression with the one given in \eref{3-25} is 
	obvious. 
	\item $v=4,r=2$:
	\bea
	C^{(2)}_{\mathbf{v}_3\to 3;\widehat{ij}}&=&\frac{(D-2) \overtr{H_i L} \overtr{H_j L}_{(i)} \overtr{S L}^3}{6 \overtr{L L}_{(i)} \overtr{L L}}-\frac{(D-2) \overtr{H_j L}_{(i)} \overtr{H_i S} \left(\overtr{S L}_{(i)}^2+\overtr{S L} \overtr{S L}_{(i)}+\overtr{S L}^2\right)}{6 \overtr{L L}_{(i)}}\nn
	&&+\frac{1}{6} (D-2) \overtr{H_i S} \overtr{H_j S}_{(i)} \left(\overtr{S L}_{(i)}+\overtr{S L}+\overtr{S L}_{(ij)}\right)+( i \leftrightarrow j )\co\nn
	C^{(2)}_{\mathbf{v}_3\to 3;\widehat{i}}&=&\frac{\overtr{H_i L} \overtr{S L} \left((D-2) \overtr{S L}^2+3 \overtr{L L}_{(i)} \left(R^2-\overtr{S S}\right)\right)}{3 \overtr{L L}_{(i)} \overtr{L L}}+\frac{(D-1) \overtr{H_i L} \overtr{S L}^3}{3 \overtr{L L}^2}\nn
	&&+\frac{\overtr{H_i S} \left(\overtr{L L}_{(i)} \left(\overtr{S S}_{(i)}+2 \overtr{S S}-3 R^2\right)-(D-2) \left(\overtr{S L}_{(i)}^2+\overtr{S L} \overtr{S L}_{(i)}+\overtr{S L}^2\right)\right)}{3 \overtr{L L}_{(i)}}\co\nn
	C^{(2)}_{\mathbf{v}_3\to 3}&=&\frac{2 (D-1) \overtr{S L}^3}{3 \overtr{L L}^2}+\frac{2 \overtr{S L} \left(R^2-\overtr{S S}\right)}{\overtr{L L}}\ed
	\eea
	Here, when calculating the reduction coefficients, we take the coefficient of $z^{\mathbf{v}_3-1}$ for terms contains one $R^2$ and one $S$ while we take the coefficient  of $t^2z^{\mathbf{v}_3-1}$ for terms contains three $S$'s.
\end{itemize}
\subsection{Boxes}

For the box topology, we present three cases : (1) $r=1,v-n=0$, (2) $r=0,v-n=1$, (3) $r=1,v-n=1$. As pointed out in the triangle topology, the reduction coefficients have some similarity between different topology. In fact, the similarity is classified by the pair $(r,v-n)$ as one can check by using the results listed in the Appendix and the main body of the paper.  
\begin{itemize}
	\item $r=1,v-n=0$:
	\bea
	\tensorc{1}{4}{i}&=&\frac{\overtr{H_i L} \overtr{V L}}{\overtr{L L}}-\overtr{H_i V},\nn
	\tensorselfc{1}{4}&=&\frac{2 \overtr{V L}}{\overtr{L L}}\ed
	\eea
	\item $r=0,v-n=1$:
	\bea
	C_{\mathbf{v}_4\to 4;\widehat{ijk}}&=&-\frac{1}{16} (D-2) \overtr{L L}_{(ijk)} \left(\overtr{H_j L}_{(i)} \overtr{H_k L}_{(ij)} \overtr{H_i Z}+\text{permutations}\ \text{of}\ (ijk)\right)\co\nn
	C_{\mathbf{v}_4\to 4;\widehat{ij}}&=&\frac{1}{8} (D-3) \overtr{L L}_{(ij)} \left(\overtr{H_j L}_{(i)} \overtr{H_i Z}+(i\leftrightarrow j)\right)\co\nn
	C_{\mathbf{v}_4\to 4;\widehat{i}}&=&-\frac{1}{4} (D-4) \overtr{L L}_{(i)} \overtr{H_i Z}\co\nn
	C_{\mathbf{v}_4\to 4}&=&\frac{1}{2} (D-5) \overtr{Z L}\ed
	\eea
	\item $r=1,v-n=1$:
	\allowdisplaybreaks
	\bea
	C^{(1)}_{\mathbf{v}_4\to 4;\widehat{ijk}}&=&-\frac{1}{16} (D-2) \overtr{L L}_{(ijk)} \left(\overtr{H_k L}_{(ij)} \overtr{H_i S} \overtr{H_j S}_{(i)}+\text{permutations}\ \text{of}\ (ijk)\right)\co\nn
	C^{(1)}_{\mathbf{v}_4\to 4;\widehat{ij}}&=&\frac{1}{8} (D-3) \overtr{L L}_{(ij)} \left(\overtr{H_i S} \overtr{H_j S}_{(i)}+(i\leftrightarrow j)\right)\co\nn
	C^{(1)}_{\mathbf{v}_4\to 4;\widehat{i}}&=&-\frac{(D-4) \left(\overtr{L L} \overtr{H_i S} \left(\overtr{S L}_{(i)}+\overtr{S L}\right)-\overtr{H_i L} \overtr{S L}^2\right)}{4 \overtr{L L}}\co\nn
	C^{(1)}_{\mathbf{v}_4\to 4}&=&\frac{1}{2} \left(\frac{(D-4) \overtr{S L}^2}{\overtr{L L}}-\overtr{S S}\right)\ed
	\eea

\end{itemize}

\subsection{Pentagons}

For pentagon topology, the reduction coefficients are similar, so we just present one example, i.e., scalar pentagon with higher poles. 

\begin{itemize}
	\item $v=6$
	\bea
	C_{\mathbf{v}_5\to 5;\widehat{ijkl}}&=&\frac{1}{32} (D-2) \overtr{L L}_{(ijkl)} \Big(\overtr{H_i L}_{(jk)} \overtr{H_j Z} \overtr{H_k L}_{(j)} \overtr{H_l L}_{(ijk)}+\text{permutations}\ \text{of}\ (ijkl)\Big)\co\nn
	C_{\mathbf{v}_5\to 5;\widehat{ijk}}&=&-\frac{1}{16} (D-3) \overtr{L L}_{(ijk)} \Big(\overtr{H_i L}_{(j)} \overtr{H_k L}_{(ij)} \overtr{H_j Z}+\text{permutations}\ \text{of}\ (ijk)\Big)\co\nn
	C^{(1)}_{5\to 5;\widehat{ij}}&=&\frac{1}{8} (D-4) \overtr{L L}_{(ij)} \left(\overtr{H_j L}_{(i)} \overtr{H_i Z}+(i\leftrightarrow j)\right)\co\nn
	C_{\mathbf{v}_5\to 5;\widehat{i}}&=&-\frac{1}{4} (D-5) \overtr{L L}_{(i)} \overtr{H_i Z}\co\nn
	C_{\mathbf{v}_5\to 5}&=&\frac{1}{2} (D-6) \overtr{Z L}\ed
	\eea
	\end{itemize}
    One thing we want to point out is that these coefficients have a manifest permutation symmetry. Using these observations, the expression can be very compact, as shown above. 
    \section{General expression of reduction coefficients}
		\label{apdix:generalcoeff}
	One can solve the recursion relations \eref{Enk-ViRecursion} in section \ref{sec:one-loop reduction} iteratively\footnote{For the degenerate case discussed in the section \ref{sec:deQ-reduction}, one can do similar computation, although it will be more complicated.}  and get general expressions for the final reduction coefficients. Here we present only the final results for non-degenerate $Q$ without derivation details.
	We define 
	\bea
	\mathscr{F}^{n,k,i}_{i-n_v,s}=\mathscr{P}^{k,i}_{i-n_v+1,s}\prod_{l=0}^{s}\a_{n,k-2l}
	\eea
	with
	\bea
	\mathscr{P}^{k,i}_{i-d,s}=(i+1-d)\mathscr{P}^{k,i}_{i-d+2,s-1}+(k-i+1+d-2s)\mathscr{P}^{k,i}_{i-d+1,s-1}\co~~~\label{C-2}
	\eea
	where we set $\mathscr{P}^{k,i}_{i-d,s}=0$ for either $d<0$ or $s<0$. With the initial condition $\mathscr{P}^{k,i}_{0,0}=1$, \eref{C-2} can be solved. For simplicity, we define
	\bea
	&&E_{n-l,k-l}^{(\mathbf{b}_l)}=E_{n-l,k-l}^{(b_1,b_2,b_3,\ldots,b_l)},~~~~~
	\overtr{WV}_{(\mathbf{b}_l)}=\overtr{WV}_{(b_1,b_2,b_3,\ldots,b_l)}\nn
	&&\overtr{HL}_{\what{\mathbf{b}_l}}=\overtr{H_{b_1}L}\overtr{H_{b_2}L}_{(b_1)}\overtr{H_{b_3}L}_{(b_1b_2)}\cdots \overtr{H_{b_l}L}_{(\mathbf{b}_{l-1})}\nn
	&&\overtr{LL}^{(s;\mathbf{j}_l)}_{({\mathbf{b}_l})}=\overtr{LL}^{j_1}\overtr{LL}_{(b_1)}^{j_2}\overtr{LL}_{(b_1,b_2)}^{j_3}\cdots\overtr{LL}_{(\mathbf{b}_{l-1})}^{j_l} \overtr{LL}_{(\mathbf{b}_{l})}^{s-j_1-j_2-j_3-\cdots-j_l}
	\eea
	and
	\bea
	\sum_{\{\mathbf{l,j,s}\}_a}^i&=&\sum_{l_1=0}^{i-1}\sum_{j_1=0}^{l_1}\sum^{i-l_1-1}_{s_1=\lceil{i-l_1-1\over 2}\rceil}\sum_{l_2=0}^{l_1-1}\sum_{j_2=0}^{l_2}\sum^{l_1-l_2-1}_{s_2=\lceil{l_1-l_2-1\over 2}\rceil}\cdots \sum_{l_a=0}^{l_{a-1}-1}\sum_{j_a=0}^{l_a}\sum^{l_{a-1}-l_a-1}_{s_a=\lceil{l_{a-1}-l_a-1\over 2}\rceil}\nn
	\mathcal{F}^{n,k,i}_{\{\mathbf{l,s}\}_a}&=&\mathscr{F}^{n,k,i}_{l_1,s_1}\mathscr{F}^{n-1,k-1-2s_1,l_1}_{l_2,s_2}\mathscr{F}^{n-2,k-2-2s_1-2s_2,l_2}_{l_3,s_3}\cdots \mathscr{F}^{n-a,k-a-2s_1-2s_2-\ldots-2s_{a-1},l_{a-1}}_{l_a,s_a}\nn
	\mathcal{S}_{l_0,\{\mathbf{l,j,s}\}_a;\mathbf{b}_a}&=&\prod_{p=1}^a\overtr{SS}_{(\mathbf{b}_{p-1})}^{-1 + l_{p-1} - l_p - s_p}\overtr{SL}_{(\mathbf{b}_{p-1})}^{1-l_{p-1}+l_p+2s_p}\overtr{SS}_{(\mathbf{b}_a)}^{l_a-j_a\over 2}\overtr{SL}_{(\mathbf{b}_a)}^{j_a}\co
	\eea
	where specially
	\bea
	&&\sum_{\{\mathbf{l,j,s}\}_0}^i=\sum_{j_0=0}^i,~~~~~\mathcal{F}^{n,k,i}_{\{\mathbf{l,s}\}_0}=1,~~~~~\overtr{HS}_{\what{\mathbf{b}_0}}=1\co\nn
	&&\mathcal{S}_{i,\{\mathbf{l,j,s}\}_0;\mathbf{b}_0}=\overtr{SS}^{{i-j_0\over 2}}\overtr{SL}^{j_0},~~~~~\mathscr{E}_{n,k-2\abs{\mathbf{s}_0},l_0;j_0}=\mathscr{E}_{n,k,i;j_0}\ed
	\eea
	Then we denote some notations used in dimension shifting as follows
	\bea
	\mathscr{K}^-_{n,k,s}&=&{1\over \prod_{l=1}^{s} (\b_{n,k-2s+2l})},~~~~~\mathscr{K}^-_{n,k,s;j}={-\a_{n,k-2s+2j}\over \prod_{p=1}^{j}\b_{n,k-2s+2p}}\co\nn
	\mathscr{K}^+_{n,k,s}&=&\prod_{p=1}^s\b_{n,k+2s+2-2p},~~~~~~\mathscr{K}^+_{n,k,s;j}=\a_{n,k+2s-2j}\prod_{p=1}^j\b_{n,k+2s+2-2p}\co
	\eea
	and
	\bea
	&&\mathcal{K}^{-;\mathbf{b}_l}_{n,k,2s;l}=\sum_{j_1=1}^s\sum_{j_2=1}^{s-j_1}\sum_{j_3=1}^{s-j_1-j_2}\cdots \sum_{j_l=1}^{s-j_1-j_2-\ldots-j_{l-1}}{1\over \overtr{LL}^{(s;\mathbf{j}_l)}_{(\mathbf{b}_l)} }\times\nn
	&& {\mathscr{K}^-_{n,k,s;j_1}\mathscr{K}^-_{n-1,k-1,s-j_1;j_2}\mathscr{K}^-_{n-2,k-2,s-j_1-j_2;j_3}\cdots \mathscr{K}^-_{n-l,k-l,s-j_1-j_2-j_3-\cdots-j_l}}\nn
	&&\mathcal{K}^{+;\mathbf{b}_l}_{n,k,2s;l}=\sum_{j_1=0}^s\sum_{j_2=0}^{s-j_1}\sum_{j_3=0}^{s-j_1-j_2}\cdots \sum_{j_l=0}^{s-j_1-j_2-\ldots-j_{l-1}}\times\nn
	&& {\mathscr{K}^+_{n,k,s;j_1}\mathscr{K}^+_{n-1,k-1,s-j_1;j_2}\mathscr{K}^+_{n-2,k-2,s-j_1-j_2;j_3}\cdots \mathscr{K}^+_{n-l,k-l,s-j_1-j_2-j_3-\cdots-j_l}}\nn
	&&\mathcal{K}^{\mathbf{b}_l}_{n,k,2s;l}=\mathcal{K}^{Sign(s);\mathbf{b}_l}_{n,k,\abs{2s};l}\ed
	\eea
	The final expression for general reduction coefficients is 
    \bea
    C^{(r)}_{\mathbf{v}_n\to n;\what{\mathbf{b}_{a}}}
    &=&\sum_{a_1+a_2=a,\atop \sigma\in S[\mathbf{b}_a]}\sigma\Bigg[\sum_{i=0}^{r}\sum_{\{\mathbf{l,j,s}\}_{a_1}}^{\xi(n,v,i)}c^{n,v,r}_{i,a_1,a_2,\{\mathbf{l,j,s}\}_{a_1}}(R^2)^{{r-i\over2}}\times\nn
    &&\mathcal{S}_{\xi,\{\mathbf{l,j,s}\}_{a_1};\mathbf{b}_{a_1}}\overtr{HS}_{\what{\mathbf{b}_{a_1}}}\overtr{HL}_{(\mathbf{b}_{a_1})\what{\mathbf{b}_{a_2}}}\mathcal{K}^{(\mathbf{b}_{a_1})\mathbf{b}_{a_2}}_{n-a_1,{-\mu-2\abs{\mathbf{s}_{a_1}}-l_{a_1}-j_{a_1}};a_2}\Bigg]\Bigg\vert_{t^iz^{\mathbf{v}_n-1}}\co
    \eea
    where we have defined  $\mu(n,v,r,i)=r-i+2n-2v,\xi(n,v,i)=v-n+i$, and the dimensionless factor is
	\bea
	c^{n,v,r}_{i,a_1,a_2,\{\mathbf{l,j,s}\}_{a_1}}&=&{(-)^{r+v+n+a_1+a_2}\Gamma(v-D/2-r)i!\over \Gamma(n-a_1-a_2-D/2)(v-n+i)!}\times \nn
	&&\mathcal{F}^{n,n-D-\mu(n,v,r,i),\xi(n,v,i)}_{\{\mathbf{l,s}\}_{a_1}}\mathscr{E}_{n-a_1,n-a_1-D-\mu(n,v,r,i)-2\abs{\mathbf{s}_{a_1}},l_{a_1};j_{a_1}}\ed
	\eea
	%


\bibliographystyle{JHEP}

\bibliography{reference}
	
\end{document}